\documentclass[range]{ar2e}

\usepackage{rotating}%

\usepackage{subcaption}

\usepackage{amsmath}
\usepackage{bm}
\newcommand{\fet}[1]{\mbox{\boldmath $#1$}}

\usepackage{graphicx}
\jname{Annu. Rev. Nucl. Part. Sci.}
\jyear{2015}
\begin{document}

% Page heads
\markboth{S. Gandolfi}{Neutron Matter from Low to High Density}

% Title
\title{Neutron Matter from Low to High Density}

% Author/affiliation
\author{Stefano Gandolfi,$^1$  Alexandros Gezerlis,$^2$ J. Carlson$^1$
\vspace{1cm}
\affiliation{$^1$Theoretical Division, Los Alamos National Laboratory, Los Alamos, NM 87545, USA; email: carlson@lanl.gov, stefano@lanl.gov} 
\affiliation{$^2$Department of Physics, University of Guelph, Guelph, ON, N1G 2W1, Canada; email: gezerlis@uoguelph.ca}}

% Abstract
\begin{abstract}
Neutron matter is an intriguing nuclear system with multiple connections 
to other areas of physics. Considerable progress has been made over 
the last two decades in exploring the properties of pure neutron fluids. 
Here we begin by reviewing work done to explore the behavior of very low density
neutron matter, which forms a strongly paired superfluid and is thus similar to 
cold Fermi atoms, though at energy scales differing by many orders of magnitude.
We then increase the density, discussing work that ties the study of neutron
matter with the determination of the properties of neutron-rich nuclei and  
neutron-star crusts. After this, we review the impact neutron matter at even 
higher densities has on the 
mass-radius relation of neutron stars, thereby making contact with astrophysical
observations.
\end{abstract}

% Keywords

\maketitle

% to generate article TOC
\tableofcontents

\section{INTRODUCTION}
  The properties of neutron matter have long been recognized as critical
to the properties of neutron-rich nuclei and neutron stars.
Low-density neutron matter critically impacts 
our understanding of neutron-rich nuclei.  Similarly the equation of state
of high-density low-temperature matter is critical in determining
the properties of neutron stars, including the mass-radius relationship
and the neutron-star maximum mass.

   During the past few years interest in neutron matter have been
resurgent for several reasons.  At very low densities, neutron matter
is very similar to cold Fermi atoms near unitarity (infinite 
scattering length), since two free neutrons are very nearly bound.
This enables stringent tests of theories of fermions in this strongly-interacting regime~\cite{Giorgini:2008}.  
The equation of state~\cite{Ku:2012,Carlson:2011} and 
pairing gap of unitary fermions~\cite{Schirotzek:2008,Carlson:2005,Carlson:2008}
have been calculated precisely and measured very accurately. 
These experiments  provided severe
tests of the theories; some of the calculations (including
those reviewed here) provided excellent predictions while others were
less successful.

   There has also been a resurgence of interest in the properties
of neutron-rich nuclei.  New facilities, such as FRIB, will probe the properties of many hitherto inaccessible nuclei. Among the several expected discoveries, the history of nucleosynthesis could be mentioned: we still do not know where the heavy elements were created.~\cite{Balantekin:2014} Such experiments will probe the extreme isospin-imbalance 
limit, thereby informing nuclear models and ab initio approaches. Typically
density functionals~\cite{bender2003self} are used to predict the properties of these nuclei;
while some components of the density functional are very tightly constrained
from stable nuclei the extreme isospin limit is less constrained. 
Many-body calculations of inhomogeneous neutron matter can provide
further constraints on the parameters entering density functionals.  This is also important
for the inner crust of neutron stars, where inhomogeneous neutron matter
fills the space between the lattice of heavy neutron-rich nuclei.

    Finally, neutron stars have become even more important in recent
years. The discovery of the first two-solar-mass neutron 
stars~\cite{Demorest:2010,Antoniadis:2013}
provided critical constraints on the dense matter equation of state.
These observations eliminated whole classes of models. Combined
with the recent observations of massive neutron stars, the attractive
nature of the neutron-neutron interaction at low momenta means
the equation of state must be soft at low density with a rapid transition
to a high-pressure when the higher-momentum neutron-neutron and three-neutron
interactions become important.  

   Astrophysical observations are providing better constraints on the
whole mass-radius relation of dense matter, which is critical to a 
real understanding of neutron stars~\cite{steiner2013neutron,
steiner2012connecting,Ozel:2010,Guillot:2013}.  In addition,
neutron-star mergers are increasingly seen as an important site for
heavy-element synthesis, providing even more important reasons
for understanding the properties of neutron-rich matter.  In the near
future, we can look forward to the observation of gravitational waves
from these neutron-star mergers~\cite{bauswein2012measuring}, providing the most direct evidence of
the structure of neutron matter and neutron stars.

% Head 2
\section{VERY LOW DENSITIES: NEUTRON MATTER AND COLD ATOMS}

At very low densities the equation of state of neutron matter
is determined by the s-wave (spin-0, isospin-1) neutron-neutron interaction.
This interaction is very attractive, almost large enough to produce a bound
di-neutron.  The neutron-neutron scattering length is -18.5 
fm~\cite{Howell:1998,Trotter:1999}, much larger than typical nuclear scales;
the neutron-neutron effective range is about 2.7 fm. 

 Bertsch had proposed a model of low-density neutron matter that was simply
a zero-range interaction tuned to infinite scattering length \cite{bertsch1998},
now called the unitary limit in cold atom systems.  
Remarkably, experiments in cold atom systems shortly thereafter
provided constraints on the properties of the unitary Fermi gas (and therefore
also indirectly of low-density neutron matter).
The relevant parameter in these strongly interacting low-density
systems is the Fermi momentum $k_F = (3\pi^2 \rho)^{1/3}$, where $\rho$ is
the number density, 
times the scattering length $a$; 
in the limit that the scattering length goes to infinity all properties
of the system are fundamental constants times the corresponding free Fermi 
gas quantity.
The equation of state in this limit is characterized
by the Bertsch parameter $\xi$:
\begin{equation}
E  \ = \ \xi \ E_{FG} \ = \ \xi \ \frac{3}{5} \ \frac{\hbar^2}{2m} \ k_F^2,
\end{equation}
where $E_{FG}$ is the energy of a two-component free Fermi Gas at
number density $\rho$.
Similarly the superfluid pairing gap is characterized by the ratio  $\delta$
of the gap in the unitary limit to the Fermi energy of the non-interacting
system:
\begin{equation}
\Delta \ = \ E(N+1) - \frac{1}{2} [ E(N) + E(N+2)] \ = \ \ \delta \ E_F \ = \ \delta \ \frac{\hbar^2}{2m} \ k_F^2.
\end{equation}

Several properties of the unitary Fermi gas have been studied both theoretically and experimentally. This is a very strongly correlated system, with a pairing
gap of the order of the Fermi energy ($\delta \approx 1$), 
unlike traditional superconductors (where $\delta \approx 10^{-4}$).
We concentrate in this section on comparisons of the equation of state and superfluid pairing gaps of cold atoms and neutron matter.  While neutron matter is much more complicated, with a significant effective range and
spin-dependent interactions that give small but finite corrections at low density, important similarities remain between 
the two systems.
A detailed review of pairing in neutron matter can be found in \cite{Gezerlis:2014b},
while comparisons
to cold atom physics can be found in \cite{Gezerlis:2008,Carlson:2012}.

\subsection{Interaction and Equation of State}
The equation of state is in a sense the most fundamental property
of neutron matter.  At low densities, as we have described, it can
be related to the properties of cold atoms, which have a tunable scattering
length and an effective range that is nearly zero compared to the average
interparticle spacing.  Neutron matter, in contrast, has a fixed 
scattering length and effective range.  They can be compared to cold
atoms by comparing systems at the same scattering length times Fermi wave
number, $k_F a$, as well as the same effective
range times Fermi wave number, $k_F r_e$, (i.e. by using 
dimensionless parameters characterizing the low-density system).

The equation of state for neutron matter and cold atoms have been compared 
in Quantum Monte Carlo calculations~\cite{Gezerlis:2008,Gezerlis:2010}.
For cold atoms we assume a nearly zero-range interaction tuned to give
the same scattering length times Fermi momentum  as neutron matter.
The details of the interaction are not important as long as we are at 
low density.

For neutron matter various simple interactions have been considered. 
The simplest is a pure s-wave interaction that acts only in relative s-waves,
as these are the dominant interactions at low density (see the discussion
in Section 3.1 below).  The potential
is adjusted to have the same scattering length and effective range as
that inferred from the neutron-neutron interaction (which is, in general, 
active in higher partial waves).  We also consider
a more realistic interaction that contains both s- and p-wave interactions
\cite{Wiringa:2002} as
the latter start to become relevant at higher densities.  These
simple interactions include all the physics required for the study of low-density matter, but at higher densities (and hence higher Fermi momenta)
the full neutron-neutron and three-neutron interactions must be considered.

At low density neutron matter is a very strongly paired superfluid, requiring
the inclusion of pairing into a realistic treatment of even the equation of state.  These calculations calculate the ground state through Quantum Monte
Carlo methods, specifically Diffusion Monte Carlo. They project out
the ground state through:
\begin{equation}
| \Psi_0 \rangle \ =  \ \exp ( - H \tau ) \Psi_T \ = \ \prod_{i=1}^N \exp \left( -H \ \frac{\tau}{N} \right)\  | \Psi_T \rangle,
\end{equation}
where we take the limit of large $\tau$ to reach the ground state and
$\Psi_T$ is an approximate trial wave function for the system.
In this case we take the trial wave function to be of a BCS-Jastrow form:
\begin{equation}
| \Psi_T \rangle \ = \ \left[ \prod_{i<j} f (r_{ij})\right] \ {\cal A} \left[ 
\prod_{i,j} \phi (r_{ij}) \right],
\end{equation}
where $\phi (r)$ is an s-wave pairing function describing the pairing
of spin up and spin down neutrons, and the Jastrow function $f$ can be
used to adjust the short-range behavior of the wave function independently
of the superfluid pairing. For the appropriate choice of $\phi(r)$, this
wave function reduces to the standard Jastrow-Fermi Gas wave function.
The radial form of the function $\phi (r)$ and $f (r)$ are determined
in variational calculations.

These calculations have a fixed-node approximation that implies they
provide variational upper bounds to the true energy.  They have proven 
to be very accuate in studies of cold atom systems, where accurate lattice
calculations without a fixed-node approximation are available~\cite{Carlson:2011}.
These calculations are also in very good agreement with cold atom experiments~\cite{Ku:2012}.

\begin{figure}[!tb]
\centering
   \begin{subfigure}{0.49\textwidth} 
     \includegraphics[scale=0.26]{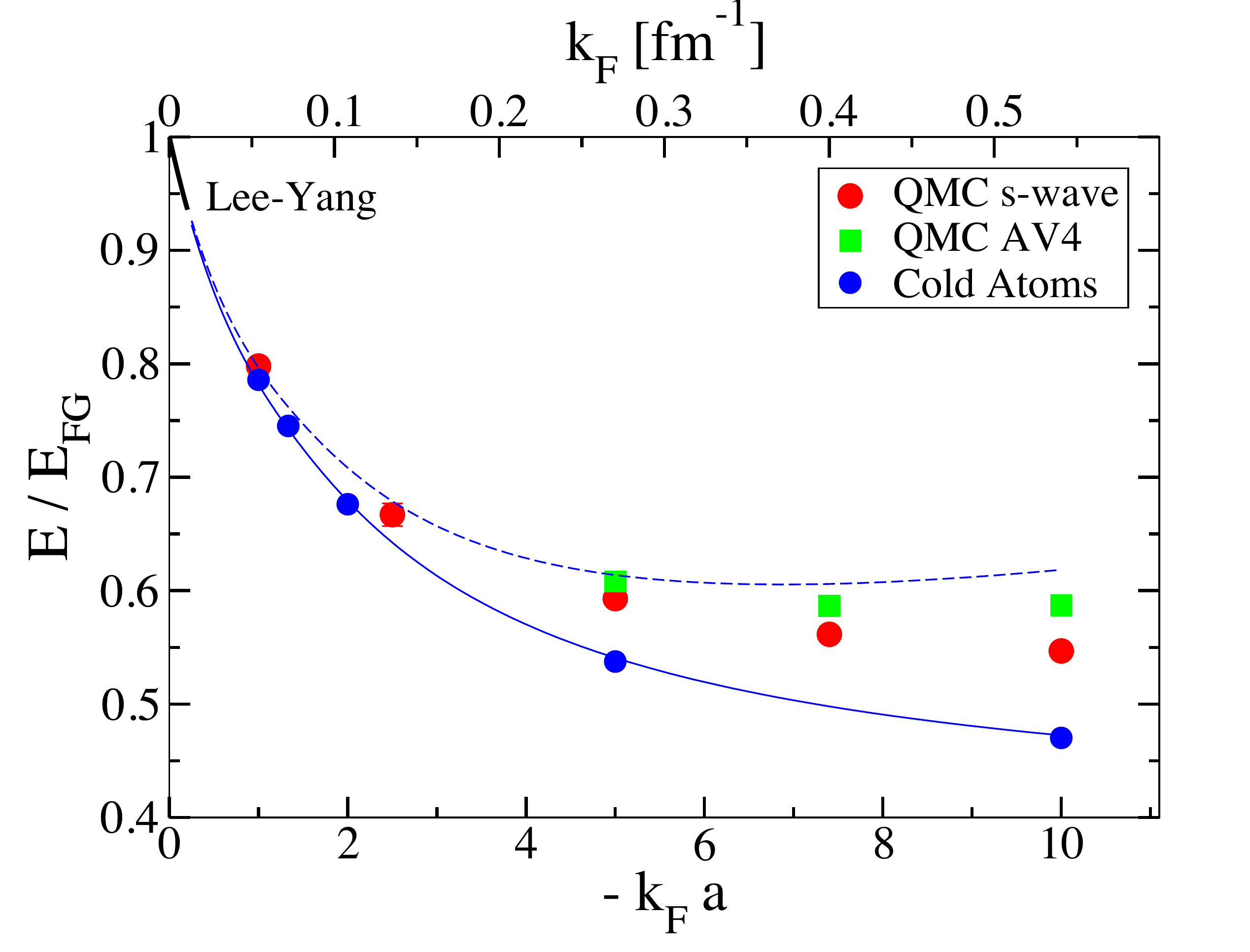}
     \caption{}
   \end{subfigure}
   \begin{subfigure}{0.49\textwidth}
     \includegraphics[scale=0.76]{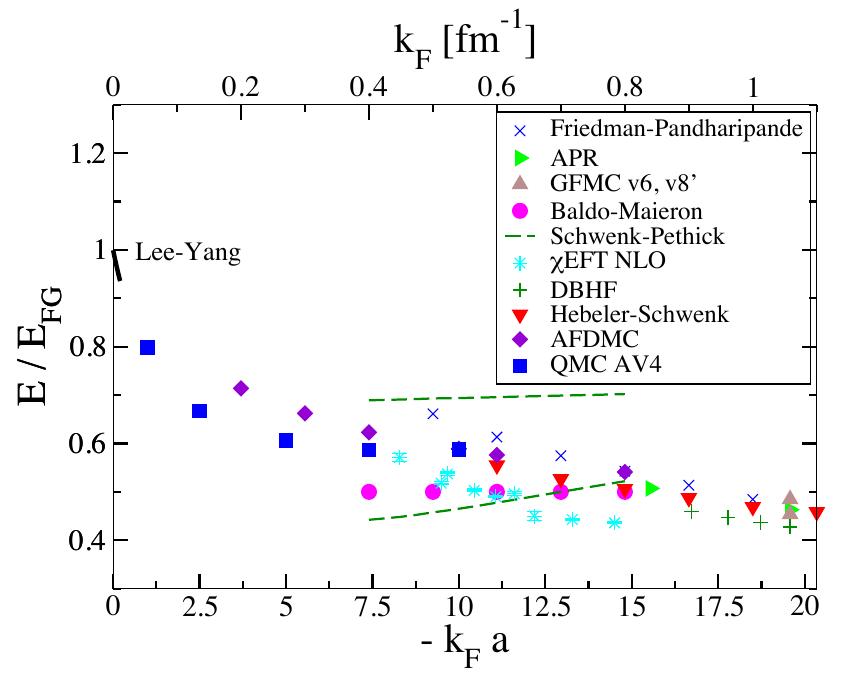}
     \caption{}
   \end{subfigure}
  \vspace{0.5cm}
\caption{ The equation of state of low-density neutron matter
compared to that of cold atoms at the same value of Fermi momentum
times scattering length ($k_F a$).  The left side compares cold atoms
and neutron matter (see text), and the right panel shows neutron matter
results for different methods over a wider range of $k_F a$.
Figures taken from \cite{Carlson:2012,Gezerlis:2010}.}
\label{fig:coldatomeos}
\end{figure}

The results of the equation-of-state calculations are shown in 
Fig. \ref{fig:coldatomeos}. The left panel compares neutron matters and
cold atoms at very low density.
The vertical axis indicates the ratio of fully interacting energy to the
energy of the free Fermi gas at the same density, the horizontal axis is
the Fermi momentum times the scattering length $k_F a$; on the upper axis
the equivalent Fermi momentum for neutron matter is indicated.
At extremely low densities, or equivalently small value of $k_F a$,
analytic results are available~\cite{Lenz:1929,Lee:1957}, and the higher-order
Lee-Yang result is plotted as a line in the figure.

Results for cold atoms with zero effective range are plotted as filled blue
circles, in the limit of infinite $k_F a$ these should approach $0.37$.  
Cold atom results for the dependence on the effective range are also available,
the equation of state can be expanded in terms of $k_F$:
\begin{equation}
E \ / E_{FG} = \xi \  +  \ {\cal S}  \ k_F  \ r_e + ... ,
\end{equation}
where ${\cal S}=0.12(3)$ is a universal constant that has been determined in the lattice
calculations and in Diffusion Monte Carlo~\cite{Carlson:2011,Forbes:2012}.
Using the above equation of including the experimental neutron-neutron effective 
range $r_e$ gives the dashed line in the figure.

The calculations of neutron matter with the spin singlet s-wave interaction
from AV18 gives the solid red points, correcting 
the p-wave interactions from AV4 gives the green squares. 
At very low densities all these calculations are very similar. 
At slightly higher
densities the correction from the p-wave interaction is slightly repulsive, 
and is in good agreement with the effective range expansion above.

The right panel shows a comparison of various methods for the neutron
matter equation of state over a somewhat wider range of densities.
Methods include Fermi Hypernetted chain resummation techniques~\cite{friedman1981hot,akmal1998equation}, 
the Brueckner-Bethe-Goldstone expansion~\cite{Baldo:2008}, 
effective field theory~\cite{Schwenk:2005} 
and several Quantum Monte Carlo methods including
Green's function Monte Carlo~\cite{Carlson:2003} and
Auxiliary field diffusion Monte Carlo~\cite{Gandolfi:2008,Gandolfi:2009}.
All the calculations are in reasonable agreement, indicating a soft
neutron matter equation of state at low density.

\subsection{Superfluid Pairing}

Neutron pairing at low density is important in both 
neutron-rich nuclei and in the crust
of neutron stars.  Pairing in nuclei and matter has been a 
long-studied topic, see a  
review by Dean and Hjorth-Jensen~\cite{Dean:2003}. For pairing at low density,
recent work in cold atom systems, both theoretical and experimental, 
has advanced our understanding of pairing in the strongly superfluid regime.
Experiments and calculations indicate that the pairing gap in neutron matter is quite substantial, reaching a peak of approximately 30 per cent of the Fermi energy.  This is the largest Fermion pairing gap known in nature, and only slightly smaller than the 45 per cent pairing found in cold atoms at 
unitarity~\cite{Schirotzek:2008,Carlson:2005,Carlson:2008}.

\begin{figure}[!tb]
\centering
   \begin{subfigure}{0.49\textwidth} 
     \includegraphics[scale=0.26]{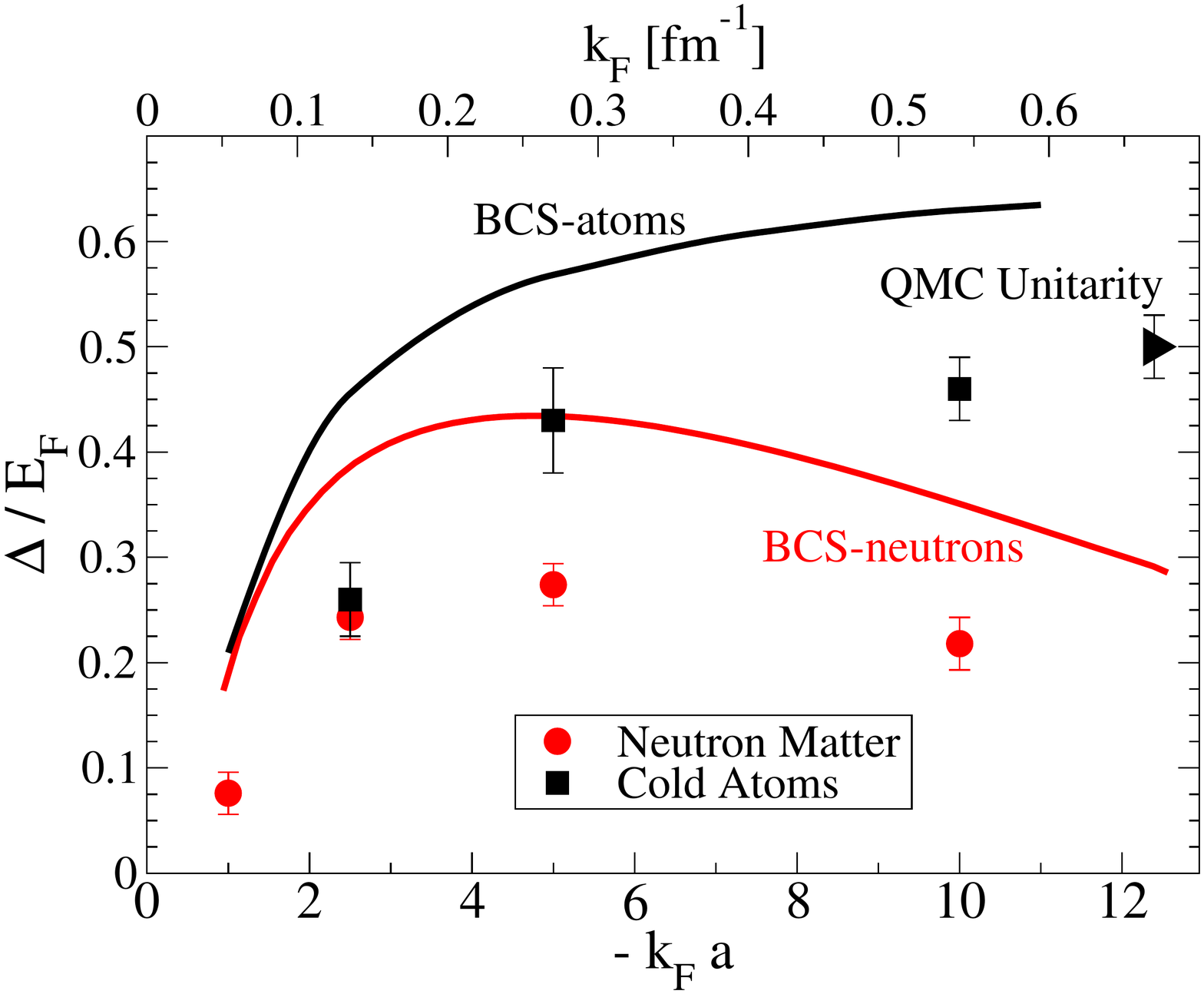}
     \caption{}
   \end{subfigure}
   \begin{subfigure}{0.49\textwidth}
     \includegraphics[scale=0.72]{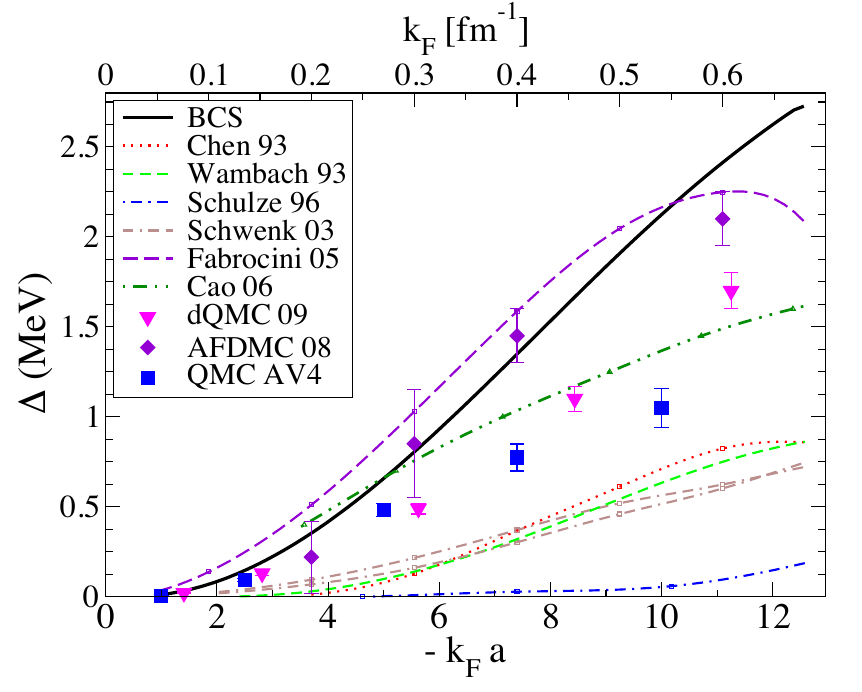}
     \caption{}
   \end{subfigure}
  \vspace{0.5cm}
\caption{ The pairing gap of low-density neutron matter
compared to that of cold atoms at the same value of Fermi momentum
times scattering length (left). Comparison of different calculations
of the neutron matter gap (right).
Figures taken from \cite{Gezerlis:2008,Gezerlis:2010}.}
\label{fig:coldatomgap}
\end{figure}

The pairing gap for matter can be calculated by computing the energy of even and odd
particle systems and extrapolating to the thermodynamic limit.  For traditional 
superfluids and superconductors this is extremely difficult as the small value of the
pairing gap implies a very large coherence length, in principle requiring extremely
large simulations with very small differences in energy.  
For this reason Quantum
Monte Carlo methods had not previously been 
used to extract superfluid pairing gaps.

Analytic techniques, are, of course available. In nuclear physics 
the BCS equations have often been used to study pairing in neutron matter.  
At low densities the s-wave pairing predictions from BCS theory are 
mostly independent of nuclear interaction
models because the interaction is very well constrained at low momenta.
However corrections to this mean-field theory varied widely, as discussed in 
~\cite{Dean:2003}.

The pairing gap of cold atoms and neutron matter are compared in 
the left panel of Fig. \ref{fig:coldatomgap}.
At very low $k_F a$  the Gorkov-Melik-Barkhudarov
analytic corrections to BCS theory are available~\cite{Gorkov:1961}.
These corrections reduce the pairing gap by a factor of $1 / (4 e)^{1/3}$
from that obtained in the simple BCS theory, resulting in a pairing gap of:
\begin{equation}
\Delta^0 (k_F) = \frac{1}{(4e)^{1/3}} \frac{8}{e^2} 
\frac{\hbar^2 k_F^2}{2m} \exp\left( \frac{\pi}{2ak_F}\right)~.
\end{equation}
This pairing gap is, however, accurate only at small values of $k_F a$,
and hence not applicable near the peak of the pairing gap.

At these larger values it is again useful to compare with the BCS-BEC
transition in cold atom systems.  In the extreme BEC limit, the BCS equation
yields a pairing gap of 1/2 the binding energy of a pair, as the 
unpaired fermion can propogate nearly freely in the medium when the
pairs are very strongly bound.  There will be no further polarization
corrections to this limit. It is natural to assume a smooth
transition between the BCS limit, where the beyond-mean-field limits
reduce the gap significantly, to the strongly-paired limit where the
corrections are very small.

This smooth transition is in fact observed in cold atom experiments
and in calculations of cold atoms and neutron matter.  The pairing
gap has been studied extensively experimentally, using both the
density harmonically trapped up and down spins in a polarized system\cite{Carlson:2008} and by studying the radio-frequency (RF) response~\cite{Shin:2008}.
Analyses of the experimental result indicate a pairing gap of $0.45 \pm 0.05
E_F$, or approximately 65 per cent of the pairing gap found in
the BCS theory.

The calculated pairing gaps in neutron matter are not as large
as those in cold atoms because of the finite effective range.
Eventually at larger momenta the s-wave interaction becomes
repulsive. These higher order corrections result in a peak
of the gap of about 0.3 times the Fermi energy at $k_F a \approx -5$,
or $k_F \approx 0.25 - 0.3$ fm$^{-1}$. 

A variety of calculations of the pairing gap are shown in the right panel
of Fig. \ref{fig:coldatomgap}. The simple BCS theory is shown as a
solid line, other calculations include correlated-basis-function
calculations~\cite{chen1993pairing,fabrocini2005s}, 
polarization-potential~\cite{wambach1993quasiparticle},
medium-polarization calculation~\cite{schulze1996medium},
Brueckner~\cite{cao2006screening},
renormalization group~\cite{schwenk2003renormalization}, 
determinantal lattice Monte Carlo calculations~\cite{abe2009low},
Diffusion Monte Carlo~\cite{Gezerlis:2010},
and AFDMC results~\cite{Gandolfi:2008,Gandolfi:2009b}.

The pairing gaps do not agree nearly as well as the equation of state.
However most of the approaches employed are not designed to handle
the strong pairing that is present for neutron matter.
Only the QMC approaches predicted the pairing
gap in cold atoms and smoothly interpolate between BCS and BEC limits.
They all result in rather large pairing gaps, with a modest reduction
of the BCS result for pairing.
The behavior of the pairing
gap at higher densities is certainly interesting and important,
particularly the p-wave gap, but requires additional study
for a reliable prediction.

\section{MODERATE DENSITIES: INTERACTIONS AND EQUATION OF STATE}
Low-density neutron matter, as seen in the previous section,
can be described by a neutron-neutron potential that is central, i.e.
a function of the interparticle spacing alone, $V(r)$. As we saw,
the essential features of the interaction can be captured by
the $s$-wave scattering length $a$ and the effective range $r_e$, i.e. only two numbers.

As the density increases different partial waves and spin-states come into the picture. 
That means that any formulation of the nucleon-nucleon potential needs to account for
all of the allowed ways in which nucleons can interact with each other. Similarly, any
quantum many-body method that is used to address interacting nucleons had better be 
able to handle the different partial waves and spin-states (and for the general case 
also isospin-states). In this section we first go over the basics of the operator
structure of standard nuclear forces (both phenomenological ones and those following
from chiral Effective Field Theory) before discussing results for the pure neutron
matter equation-of-state (energy versus density) resulting from a variety of many-body
methods.

\subsection{Neutron-Neutron Interaction}
\label{sec:eft}
\begin{figure}[!tb]
\centering
\includegraphics[width=0.6\textwidth]{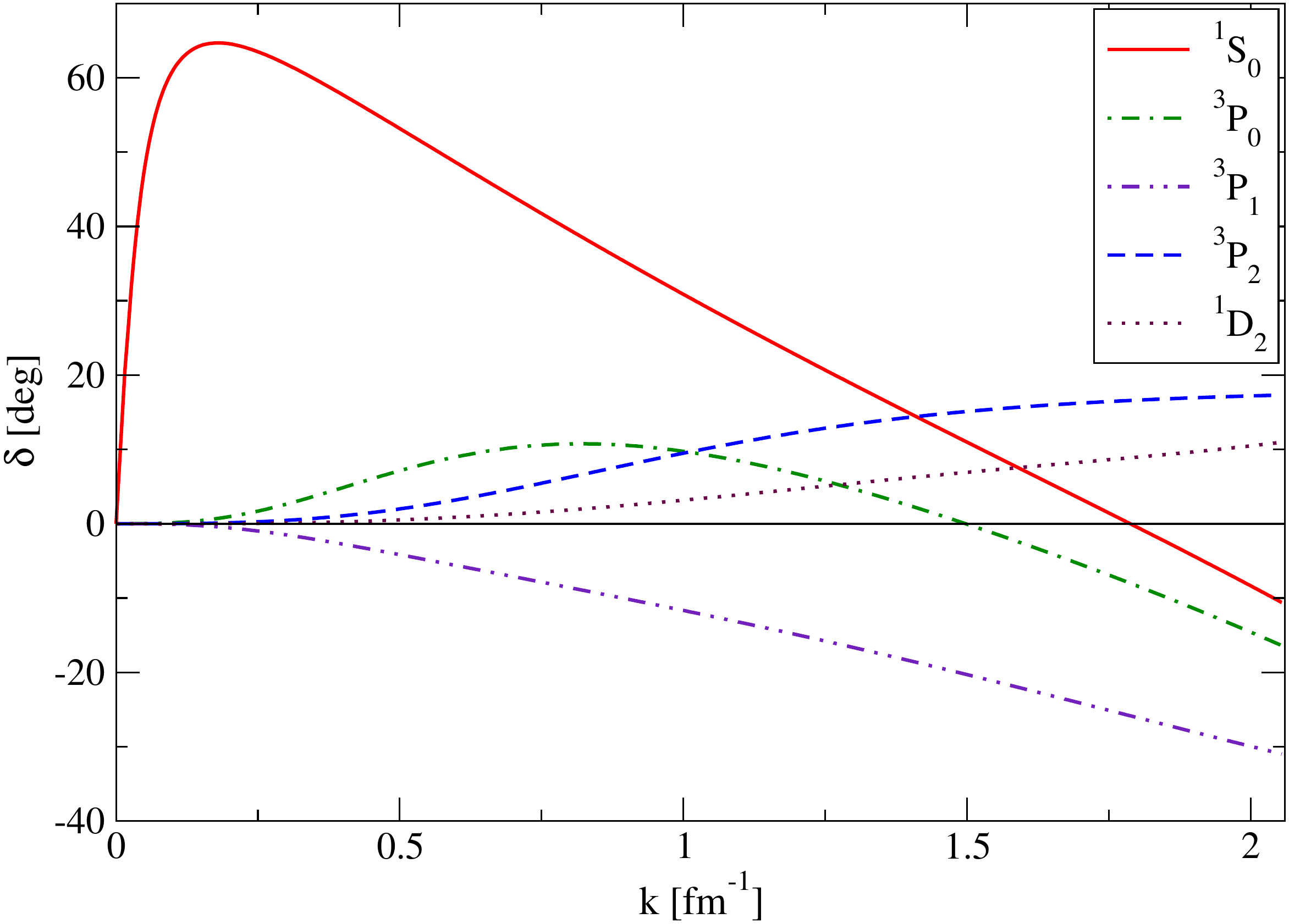}
\caption{$np$ scattering phase shifts vs the wave number of a nucleon in the center-of-mass frame 
(based on the Nijmegen93 partial-wave analysis). Charge-independence and charge-symmetry breaking
terms would slighly change these results for the case of neutron-neutron scattering.
Figure taken from \cite{Gezerlis:2014b}.}
\label{fig:phaseshifts}
\end{figure}

Nucleon-nucleon (to be precise, $np$) scattering in different channels can be experimentally probed, so any nuclear force needs to reproduce a large number of scattering phase shifts. 
The different ``channels'' in which phase shifts are extracted from experiment 
are
conveniently encapsulated in the spectroscopic notation $^{2S+1}L_J$: 
L denotes the orbital angular momentum using S,P, D, and so on. 
The $2S+1$ exponent for two neutrons can be either $2 \cdot 0 + 1 = 1$ (spin-singlet)
or $2 \cdot 1 + 1 = 3$ (spin-triplet).
Thus, the low-density $s$-wave scattering mentioned above which is dominant 
at low densities in neutron matter is given as $^1$S$_0$. This is clearly seen
in Fig.~\ref{fig:phaseshifts}, which displays scattering phase shifts vs nucleon
wave number/momentum: at vanishing momenta (corresponding to vanishing densities)
the only phase shift that is non-zero is precisely the one in the $^1$S$_0$ channel.
(the $^3$S$_1$ channel, familiar from the deuteron, does not come into play for
the neutron-neutron case).
We remind the reader that a positive phase shift corresponds to an attractive interaction: 
this implies that at very low densities the interaction in the $^1$S$_0$ channel is attractive,
turning repulsive at around $k = 1.7$ fm$^{-1}$ (thereby retroactively explaining the gap 
closure we saw in the previous section). At higher densities/momenta, $^3$P$_2$ 
is clearly the dominant channel, leading to $^3$P$_2$-$^3$F$_2$ pairing (the coupling being
due to the tensor operator, see below).

Traditionally, the effective reproduction of phase shifts 
was accomplished via the breakup of the nucleon-nucleon potential
in different channels, exemplified by the Argonne family of potentials~\cite{Wiringa:1995}.
Here is the Argonne v8' (AV8') \cite{Wiringa:2002} case:
\begin{equation}
V_{12} = \sum_{p=1}^{8} v^p(r) O^{p}_{12}\,,
\label{eq:av8p}
\end{equation}
Here $O^{p}_{12}$ are spin-isospin dependent operators, which are given by:
\begin{equation}
O^{p=1,8}_{12}=(1,\fet \sigma_1 \cdot \fet \sigma_2,S_{12},
{\bf L}\cdot {\bf S})\times(1,\fet \tau_1 \cdot \fet \tau_2)\,.
\end{equation}
In this expression $S_{12} = 3 \fet \sigma_1 \cdot \hat {\bf{r}} \; \fet
\sigma_2 \cdot \hat {\bf r} - \fet \sigma_1 \cdot \fet \sigma_2$ is
the tensor operator, where ${\bf r}={\bf r}_1-{\bf r}_2$ is 
the nucleon separation vector. 
Similarly, ${\bf L} = -i {\bf r}\times (\nabla_1-\nabla_2)$
is the relative
angular momentum 
and ${\bf S} = \fet \sigma_1 + \fet \sigma_2$ is the total spin for the pair $12$.
An $nn$ pair can only exist in an isotriplet state ($T=1$), 
so $\fet \tau_1 \cdot \fet \tau_2 = 1$. The $ v^p(r)$ functions
in Eq.~(\ref{eq:av8p}) contain one-pion exchange at large distances
and high-quality phenomenology at intermediate and short distances.

More recently, the nuclear physics communicty has embraced chiral Effective Field Theory
(EFT) interactions, in an attempt to connect with the symmetries of the underlying
fundamental theory (Quantum Chromodynamics)~\cite{Epelbaum:2009,Machleidt:2011}.
In what follows, we provide a bare-bones description of chiral EFT, 
starting with the features that are common to all modern chiral EFT potentials. 
 Chiral EFT interactions employ
 a separation of scales, that between pion and vector meson masses, and
 attempt to systematically expand in a small parameter that is the ratio
 between the two. Assuming the power counting employed is self-consistent,
 this provides a hierarchy of forces controlled by the power of the expansion.
Thus, chiral EFT systematically includes the analytically derived long-range one-pion 
exchange, as well as the intermediate-range two-pion exchanges, and so on. 
 In addition to this, the power counting naturally leads to consistent three-nucleon
 and many-nucleon forces (e.g. at leading-order, LO, and next-to-leading order, NLO, 
 no 3N forces are predicted). The soft scale, identified with the pion mass above, 
 is generally taken to be that of a relevant momentum scale. 
 In addition to the known pion exchanges, chiral EFT interactions also 
 include phenomenological short-range terms, typically written down as Dirac-delta
 functions/contact terms: these do not reflect any deeper physics other than respecting
all relevant symmetries and the power counting.

Before we look at specific expressions for the different parts of a chiral EFT interaction,
we go over some definitions. Since chiral EFT interactions are based on a low-momentum
expansion, it was natural that they were first formulated in momentum space.
If we denote the incoming and outgoing
relative momenta by
${\bf p}=({\bf p}_1-{\bf p}_2)/2$ and ${\bf p}'=({\bf p}'_1-{\bf
  p}'_2)/2$, respectively, then we can define the momentum 
 transfer ${\bf q}={\bf p}'-{\bf p}$ and the momentum transfer
 in the exchange channel ${\bf k}=({\bf p}'+{\bf p})/2$. 
 Generally speaking, terms that depend on ${\bf q}$ are local and those
 depending on ${\bf k}$ are non-local. The power expansion mentioned above 
 is therefore an expansion in powers of ${\bf q}$ and ${\bf k}$. Since our discussion 
 is meant to be somewhat pedagogical, we limit ourselves below to the first three 
 orders in the chiral expansion though, obviously, the pion exchanges and contact 
 terms have also been extensively studied at higher orders.

Let us now look at the one- and two-pion exchanges in momentum space:
\begin{align}
V^{\text{mom}}_{1\pi, LO} &= U_T(q) ~(\fet \tau_1 \cdot \fet \tau_2) ~(\fet \sigma_1 \cdot \mathbf{q}) ~(\fet \sigma_2
\cdot \mathbf{q}) \nonumber \\
V^{\text{mom}}_{2\pi, NLO} &= W_C(q) ~(\fet \tau_1 \cdot \fet \tau_2) + 
V_S(q) ~(\fet \sigma_1 \cdot \fet \sigma_2)  +  
V_T(q) ~(\fet \sigma_1 \cdot \mathbf{q}) ~(\fet \sigma_2
\cdot \mathbf{q}) \nonumber \\
V^{\text{mom}}_{2\pi, N^2LO} &= V_C(q) + 
W_S(q) ~(\fet \tau_1 \cdot \fet \tau_2) ~(\fet \sigma_1 \cdot \fet \sigma_2)  + 
W_T(q) ~(\fet \tau_1 \cdot \fet \tau_2) ~(\fet \sigma_1 \cdot \mathbf{q}) ~(\fet \sigma_2
\cdot \mathbf{q})
\end{align}
Note that all three of these expressions are functions of $q$ alone (i.e. they are not functions of $k$).
Thus, they can be trivially Fourier transformed and take the following form in coordinate space:
\begin{align}
V^{\text{coord}}_{1\pi, LO} &= U_S(r) ~(\fet \tau_1 \cdot \fet \tau_2) ~(\fet \sigma_1 \cdot \fet \sigma_2) 
+ U_T(r)  ~(\fet \tau_1 \cdot \fet \tau_2) ~S_{12} \nonumber \\
V^{\text{coord}}_{2\pi, NLO} &= W_C(r) ~(\fet \tau_1 \cdot \fet \tau_2) 
+ V_S(r) ~(\fet \sigma_1 \cdot \fet \sigma_2)
+ V_T(r)  ~S_{12} \nonumber \\
V^{\text{coord}}_{2\pi, N^2LO} &= V_C(r)  
+ W_S(r) ~(\fet \tau_1 \cdot \fet \tau_2) ~(\fet \sigma_1 \cdot \fet \sigma_2)
+ W_T(r)  ~(\fet \tau_1 \cdot \fet \tau_2) ~S_{12}
\end{align}
where we are only showing the finite-range parts of the pion exchange terms.
It is easy to see, for example by focusing on the one-pion exchange 
term, that the $(\fet \sigma_1 \cdot \mathbf{q}) ~(\fet \sigma_2
\cdot \mathbf{q})$ term in momentum space 
gives rise to $\fet \sigma_1 \cdot \fet \sigma_2$ and $S_{12}$ terms in coordinate space.

Note that the above expressions for the pion exchanges do not contain relativistic 
$1/m_N^2$ and $1/m_N$ corrections for the one- and two-pion exchange, respectively 
(where $m_N$ is the nucleon mass). These would also involve (isoscalar and isovector) 
spin-orbit terms. This means that in chiral EFT the only spin-orbit term up to order N$^2$LO appears
in the short-range contacts, which, as mentioned above, respect 
all relevant symmetries as well as the power counting. Qualitatively, working in momentum space,
this means that at LO
all 4 possible spin-isospin combinations should be present, without any momentum dependence.
At NLO, something analogous holds: there are 14 different contact interactions that all
contain a momentum-squared in some form. This can be summarized as:
\begin{align}
V^{\text{mom}}_{cont,LO} &= \alpha_1 + \alpha_2 {\bm \sigma}_1 \cdot {\bm \sigma}_2 + \alpha_3 {\bm \tau}_1 \cdot {\bm \tau}_2 + \alpha_4 {\bm \sigma}_1 \cdot {\bm \sigma}_2 \, {\bm \tau}_1 \cdot {\bm \tau}_2 \nonumber \\
V^{\text{mom}}_{cont,NLO} &= \gamma_1 \, q^2 + \gamma_2 \, q^2\, {\bm \sigma}_1 \cdot {\bm \sigma}_2 + \gamma_3 \, q^2\, {\bm \tau}_1 \cdot {\bm \tau}_2 
+ \gamma_4 \, q^2 {\bm \sigma}_1 \cdot {\bm \sigma}_2 {\bm \tau}_1 \cdot {\bm \tau}_2 \nonumber \\
&\quad + \gamma_5 \, k^2 + \gamma_6 \, k^2\, {\bm \sigma}_1 \cdot {\bm \sigma}_2 + \gamma_7 \, k^2\, {\bm \tau}_1 \cdot {\bm \tau}_2 + \gamma_8 \, k^2 {\bm \sigma}_1 \cdot {\bm \sigma}_2 {\bm \tau}_1 \cdot {\bm \tau}_2 \nonumber \\
&\quad + \gamma_9 \, ({\bm \sigma}_1 + {\bm \sigma}_2)({\bf q}\times {\bf k}) + \gamma_{10} \, ({\bm \sigma}_1 + {\bm \sigma}_2)({\bf q}\times {\bf k}) {\bm \tau}_1 \cdot {\bm \tau}_2 \nonumber \\
&\quad + \gamma_{11} ({\bm \sigma}_1 \cdot {\bf q}) ({\bm \sigma}_2 \cdot {\bf q})  + \gamma_{12} ({\bm \sigma}_1 \cdot {\bf q}) ({\bm \sigma}_2 \cdot {\bf q}) {\bm \tau}_1 \cdot {\bm \tau}_2 \nonumber \\
&\quad + \gamma_{13} ({\bm \sigma}_1 \cdot {\bf k}) ({\bm \sigma}_2 \cdot {\bf k})  + \gamma_{14} ({\bm \sigma}_1 \cdot {\bf k}) ({\bm \sigma}_2 \cdot {\bf k}) {\bm \tau}_1 \cdot {\bm \tau}_2
\end{align}
(This is in momentum space, leading to a Dirac delta $\delta(r)$ in coordinate space, hence the name contacts.)
It's obvious that this general set of contacts contains both $q$ and $k$ on an equal footing.
In reality, only 2 and 7 of these contacts at LO and NLO, respectively, are linearly independent. 
As a rough mapping between momentum space and coordinate space, we note
that the isoscalar spin-orbit term $({\bm \sigma}_1 + {\bm \sigma}_2)({\bf q}\times {\bf k})$
corresponds to ${\bf L}\cdot {\bf S}$.

All the chiral EFT terms mentioned above need to somehow be cut off (``regulated'') at high momenta in
order to avoid infinities. Thus, they are multiplied with a regulator function:
\begin{equation}
f_{\Lambda}(p',p) = \exp \left [ - \left ( \frac{p'}{\Lambda} \right )^{2n} - \left ( \frac{p}{\Lambda} \right )^{2n} \right ]
\end{equation}
where $n$ is some power and $\Lambda$ is known as the cutoff parameter, typically taken to be 400-500 MeV. 
This regulator and cutoff are used in the standard momentum-space formulations of chiral EFT by Entem-Machleidt 
(EM)~\cite{Entem:2003} and Epelbaum-Gl\"ockle-Meissner (EGM)~\cite{Epelbaum:2004}. 
In addition to this, the EGM approach also uses a spectral-function regularization on the two-pion exchange
terms given above~\cite{Epelbaum:2004},
leading to a 2nd cutoff parameter, $\tilde{\Lambda}$ which is typically taken to be larger
than $\Lambda$ (700 MeV and above). It is also significant that nuclear forces with 
explicit $\Delta$ degrees of freedom are being investigated. \cite{Krebs:2007, Piarulli:2014}

More recently, two new chiral EFT potentials have appeared: i) N$^2$LO$_{\text{opt}}$~\cite{Ekstroem:2013},
which carried out
a high-quality re-fit of the low-energy constants appearing in the contact terms, like those shown above, as
well as the pion-nucleon couplings, and ii) local chiral EFT, \cite{Gezerlis:2013,Gezerlis:2014} 
which made use of the afore-mentioned ambiguity
in selecting the contacts such that they are quasi-local to begin with. [The latter approach also made use of 
a local regulator (i.e. one that is a function of $q$ or $r$), as also adopted by a very recent 
mixed version of chiral EFT~\cite{Epelbaum:2014}, which employs a local regulator and non-local contacts.] As will be 
seen in the many-neutron results discussed in the following subsection, the N$^2$LO$_{\text{opt}}$ 
potential is very soft, a feature which is useful to many-body methods that work in momentum space. 
On the other hand, the local chiral EFT approach has varied the cutoff $\Lambda$, 
producing a family of potentials, ranging from hard to quite soft.

\subsection{Equation of State}

While the nuclear force employed is an important component in describing neutron matter,
just as important is the quantum many-body theory that is used to study the many-neutron problem. 
Details on specific results are given below, but let us first go over some general features of
nuclear many-body methods. We limit our discussion to those methods that have been 
recently reformulated for (or applied to) neutron matter.

The first large distinction we could make is between \textit{ab initio}
methods, which try to solve the many-neutron Schr\"odinger equation employing varying degrees 
and types of approximations, and phenomenological approaches, which focus on reproducing and predicting 
experimentally relevant properties without necessarily seeking 
a connection to nuclear forces~\cite{Bender:2003}.
 \textit{Ab initio} methods,
in turn, can be grouped according to another general distinction, that between perturbative and non-perturbative
methods. The prototypical perturbative method employed to describe neutron matter is Many-Body Perturbation
Theory (MBPT), an approach which attempts to sum all classes of diagrams up to a specified order in interaction lines~\cite{Hebeler:2010,Tews:2013,Coraggio:2013}.
On the other hand, non-perturbative nuclear methods can also be distinguished between 
a) resummation schemes, which 
sum a specific class of diagrams to all orders in the interaction, 
typically employing a ladder approximation, \cite{Carbone:2014,Sammarruca:2014}, 
b) Coupled Cluster theory, which works by generating $np-nh$ excitations
of a reference state, typically truncated at the doubles level \cite{Hagen:2014}, 
and c) Quantum Monte Carlo methods, in which the many-body problem is solved stochastically, \cite{Schmidt:1999,Gandolfi:2009,Gezerlis:2013,Roggero:2014,Wlazlowski:2014}. 
One could legitimately add to this list d) density functional theory, which \textit{de facto} 
includes complicated many-body correlations and is applicable throughout the nuclear chart.

Given the significance of energy-density functional approaches of heavy nuclei, it is worthwhile to examine the
interplay between ab initio calculations for neutron matter and EDF approaches.
Ref.~\cite{Brown:2000} explicitly showed that the equation-of-state (EOS) of neutron
matter can be used to eliminate as unphysical several Skyrme energy-density functionals.
As a result, the EOS of neutron matter has been, for a while now, used 
as a constraint in energy-density functional theories of nuclei~\cite{Bender:2003}.
Recently, the interplay of Skyrme functionals and pure neutron systems was exploited
in the context of the polaron/impurity problem~\cite{Forbes:2014}, and more generally~\cite{Brown:2014} -- see also section 4 of the present review.
The idea behind energy-density functional theories of nuclei is to use as input as much known physics 
as possible, in order to make predictions in regions where no experiment is possible (yet) and fully
microscopic calculations are unrealistic. The use of microscopic results for pure neutron properties, 
then, ties in with that agenda, extending the input constraints beyond experiment: it stands to reason
that neutron-rich nuclei predictions will depend critically on pure neutron constraints.  Of course,
the resulting EDFs can be applied to the context of neutron stars as well~\cite{Dutra:2012,Erler:2013},  
again addressing systems that are beyond the reach of ab initio approaches. Such density
functionals could also explicitly include a neutron-star crust equation-of-state as input.

We now go over selected results on the equation of state of pure infinite neutron matter, following from 
several quantum many-body approaches. All of these results use as input chiral EFT neutron-neutron interactions.
It's important to note that if one is limited to the NN sector alone, the chiral EFT approach has no built-in
advantage over phenomenological approaches like the Argonne family of potentials or CD-Bonn. One of the main
distinguishing features of the chiral EFT approach is that it provides guidance (through its order-by-order expansion) on 
how to consistently pick the form of the 3N interaction. Thus, the results summarized in this subsection should
be seen as the foundation of an approach that combines NN+3N forces, which become even more important at higher
densities, as discussed later in this review.

\begin{figure}[!tb]
\centering
   \begin{subfigure}{0.49\linewidth} 
     \includegraphics[scale=0.26]{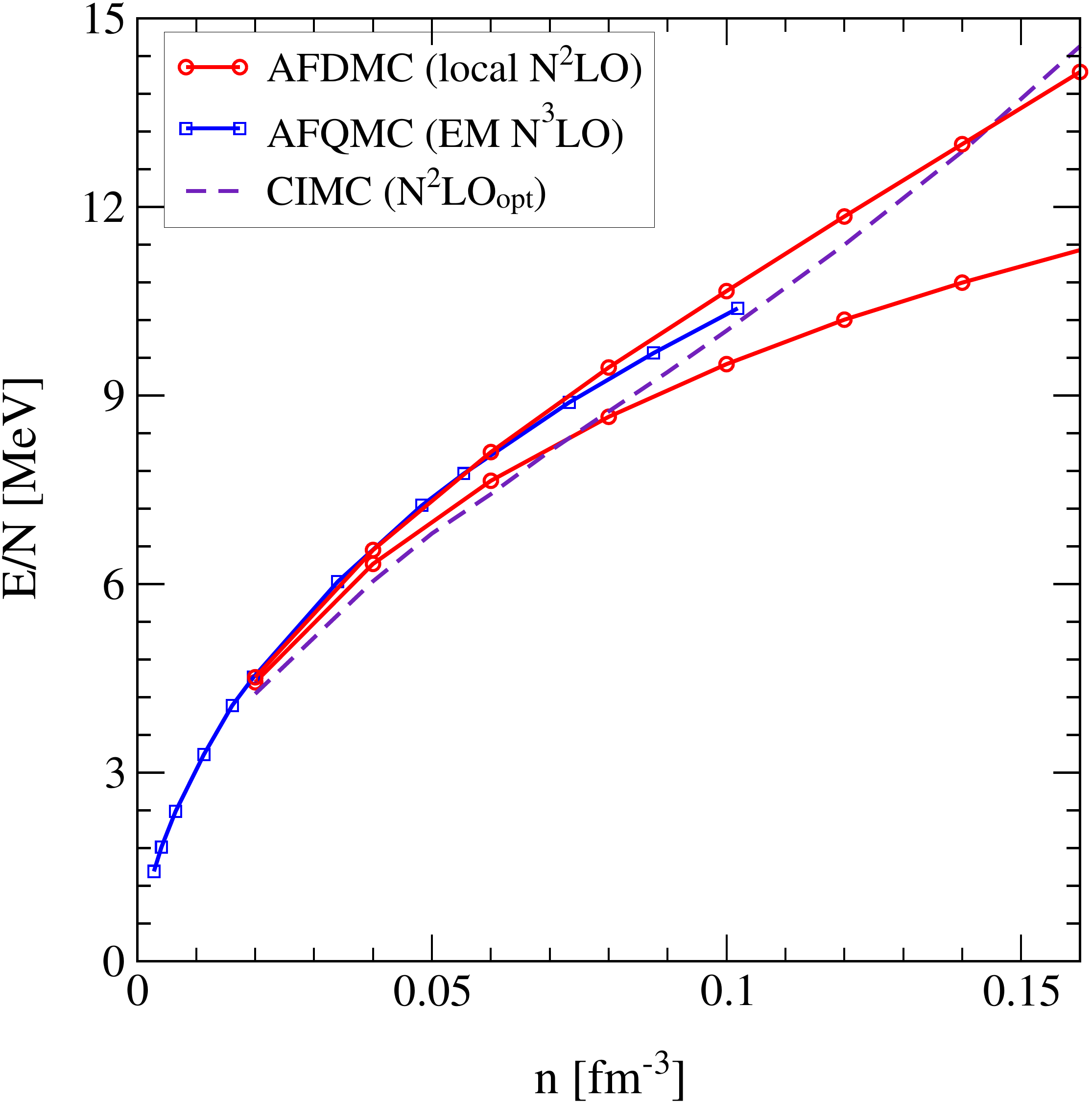}
     \caption{}
   \end{subfigure}
   \begin{subfigure}{0.49\textwidth}
     \includegraphics[scale=0.26]{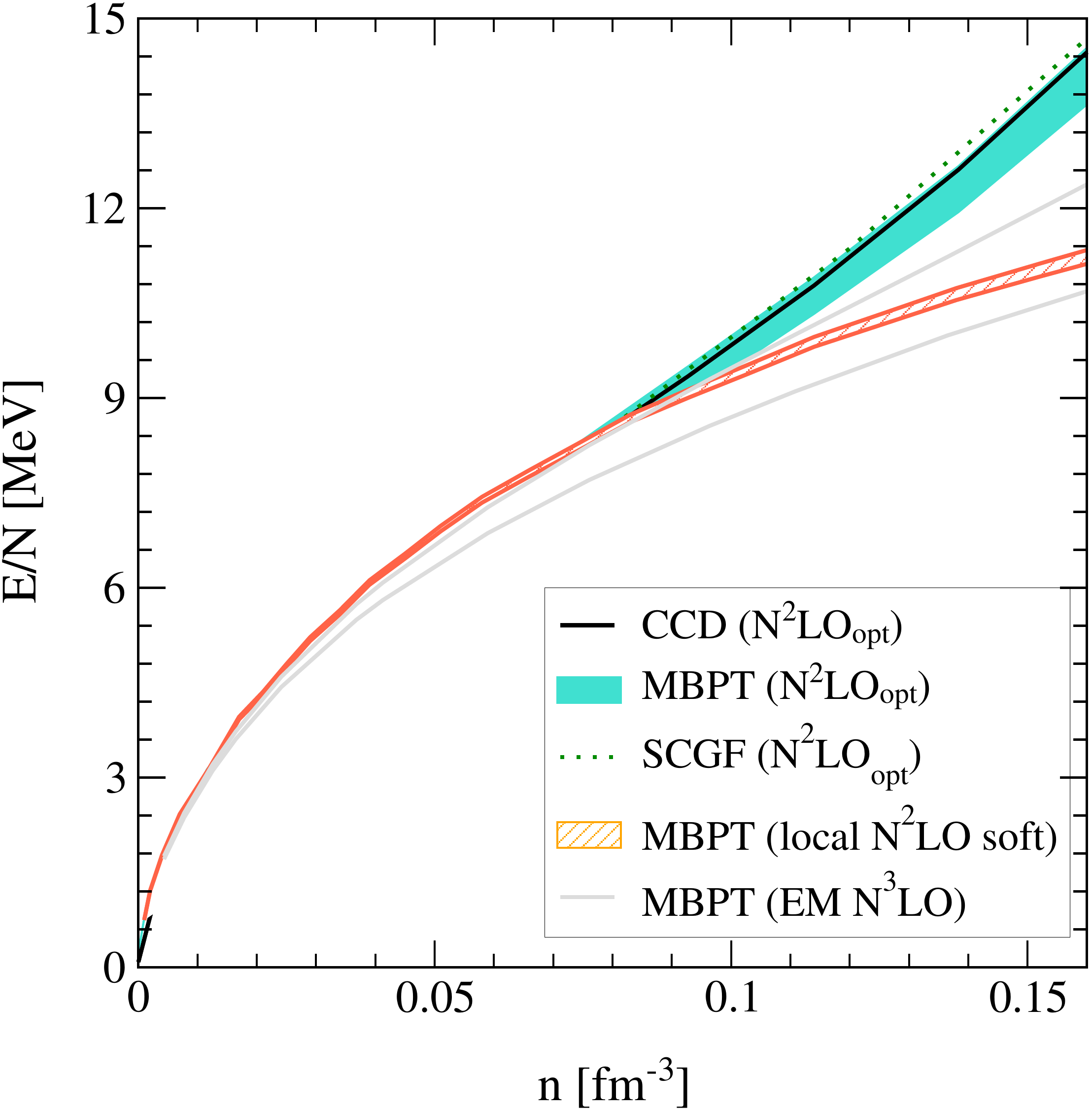}
     \caption{}
   \end{subfigure}
  \vspace{0.5cm}
\caption{Equation-of-state of intermediate density neutron matter using only NN interactions as input. Shown are results for different chiral EFT potentials, different orders in the chiral expansion, and different quantum many-body methods. (a) Quantum Monte Carlo results, (b) All other many-body results. Details are provided in the main text.}
\label{fig:eos_2b}
\end{figure}

We show in Fig.~\ref{fig:eos_2b} results for the equation of state of neutron matter, following from 
calculations that use only nucleon-nucleon interactions as input. Specifically, we show MBPT results
using local N$^2$LO interactions as input~\cite{Gezerlis:2014} (specifically the soft
400 MeV potential), MBPT results using 
an N$^3$LO EM potential~\cite{Coraggio:2013},
MBPT results using N$^2$LO$_{\text{opt}}$~\cite{Tews:2013b},
self-consistent Green's functions results using N$^2$LO$_{\text{opt}}$~\cite{Carbone:2014},
coupled-cluster doubles (CCD) results using N$^2$LO$_{\text{opt}}$~\cite{Hagen:2014},
AFDMC results using local N$^2$LO interactions~\cite{Gezerlis:2014},
Auxiliary-Field Quantum Monte Carlo (AFQMC) results using an N$^3$LO EM potential~\cite{Wlazlowski:2014},
and configuration-interaction Monte Carlo (CIMC) results using N$^2$LO$_{\text{opt}}$~\cite{Roggero:2014}.

We see that at intermediate densities essentially all many-body methods with different input potentials give basically
the same answer qualitatively (though that ceases to be the case closer to saturation density). 
At a more detailed level, within MBPT Fig.~\ref{fig:eos_2b}b seems to imply
that as we go from N$^2$LO to N$^3$LO the energy decreases. 
 Also,
the MBPT plus soft local N$^2$LO results are essentially identical to the AFDMC plus 
soft local N$^2$LO ones~\cite{Gezerlis:2014}.
Intriguingly, the 414 MeV (soft) EM potential plus AFQMC of Ref.~\cite{Wlazlowski:2014}
leads to results that are similar to the 500 MeV (hard) local chiral potential plus AFDMC
of Ref.~\cite{Gezerlis:2014}.
The MBPT results using N$^2$LO$_{\text{opt}}$ of Ref.~\cite{Tews:2013b} underline how soft/perturbative the 
N$^2$LO$_{\text{opt}}$ potential is: this implies that non-perturbative methods are not necessarily required to handle it.
A distinguishing feature of the AFDMC results of \cite{Gezerlis:2014} is the use of a 
fully non-pertubative QMC method 
to probe both soft and hard chiral potentials.
Overall, it is clear from Fig.~\ref{fig:eos_2b} that issues like the choice of different regulators and 
different cutoffs, as well as the non-locality vs locality of the potential employed, should be 
further investigated in detail.
At vanishingly small densities the effects of pairing (discussed in an earlier section) 
would probably alter this picture, but such differences would likely not be visible at this scale.

\subsection{Three Nucleon Interactions}
It is well accepted that modern nucleon-nucleon interactions cannot
describe the binding energies of nuclei with A$\geq$3, and they have to
be combined with three-body forces.  In the framework of chiral EFT, as
discussed in Sec.~\ref{sec:eft}, the appearance of three- and higher many-body forces
naturally emerges in the chiral expansion. However, in other approaches
like for the Argonne interactions, the contribution of four-body forces
is expected to be negligible. This observation comes from the fact that
in nuclei and matter the three-body potential energy is (in magnitude)
few percent than the two-body potential energy.
The Urbana-IX (UIX) potential has been introduced to give a correct 
description of few-body systems.
The UIX includes the Fujita--Miyazawa term that describes the p-wave exchange
of two pions between three nucleons, where the intermediate state has
one nucleon excited to a $\Delta$, that is eventually integrated out.
This term is the longest-range three-nucleon interaction and a very 
similar term also arises as the leading three-nucleon contribution 
(at N$^2$LO) in chiral perturbation theory~\cite{Epelbaum:2009}. 
The Urbana models also include a spin-isospin independent short-distance
term, that models the exchange of more pions between nucleons~\cite{Carlson:1981}.
The UIX force was originally proposed in combination with the Argonne AV18
and AV8$'$~\cite{Pudliner:1995}. It has been constructed to correctly
describe the saturation of nuclear matter at $\rho_0$=0.16 fm$^{-3}$,
and to reproduce the binding energy of the $^4He$.
Although this model of the three-nucleon force does not fully alleviate 
the underbinding problem in light nuclei, 
it has been extensively used to study the equation of state of nuclear
and neutron matter~\cite{Akmal:1998,Gandolfi:2009,Gandolfi:2012,Li:2008}.
It has to be noticed that the AV18+UIX Hamiltonian supported heavy
mass neutron stars years before they have been observed.

Other phenomenological models of three-body force, such as the
Illinois forces, have been obtained by fitting the binding energies
of light nuclei up to A$\le$8~\cite{Pieper:2001}, and they give an
excellent description of both ground- and excited-states of nuclei up
to $^{12}$C~\cite{Carlson:2014}.  In addition to the Fujita--Miyazawa
term the Illinois models also include s-wave two-pion-exchange and also
three-pion-exchange ring diagrams.  Unfortunately, the three-pion rings
included in the Illinois forces leads to a very strong attraction in
pure neutron-matter~\cite{Sarsa:2003,Maris:2013} at high densities,
and produces EOS too soft to be compatible with observed neutron star
masses and radii. This is a sign of important deficiencies in the model
of the three-nucleon forces, which will become especially important at
higher density and larger isospin asymmetry.  Hamiltonians based on
chiral EFT would be very valuable to address these issues. However,
such interactions should be constructed to reproduce nucleon-nucleon
scattering data up to large energies. 
A qualitative estimate is given by observing that the momentum of two 
neutrons is given by $k\approx\sqrt{E_{lab}\,m/2}$, and then if this 
is compared to the Fermi momentum of homogeneous matter we have
$\rho\approx(E_{lab}\,m/2)^{3/2}/3\pi^2$ (of course, this is only a 
back-of-the-envelope calculation, so it does not account for possible
prefactors). This clearly shows that 
if one wants to describe important correlations in neutron matter up
to large densities, then a nucleon-nucleon interaction that 
accurately fitting phase-shifts up to large $E_{lab}$ will be 
important. Of course, this raises the questions of which 
functional form for the three-nucleon force should be employed,
as well as which observables the three-nucleon parameters should
be fit to, when one interested in describing matter at high density.

\section{INHOMOGENEOUS NEUTRON MATTER}
  The neutron matter in the exterior of neutron-rich nuclei or the inner
crust of neutron stars is not homogeneous,  so it is important to study the 
properties of neutron matter in confined geometries
and with density gradients. Typically, nuclear density 
functionals have isoscalar and isovector gradient terms that 
allow for a different energy cost of density gradients
between isospin symmetric nuclear matter and pure neutron matter.
The isoscalar gradients can be very tightly constrained by the properties
of finite nuclei, but the gradients in pure neutron matter are much
less tightly constrained.  To some degree this is a result of the
fact that we have direct experimental extractions of the
charge density through electron scattering, but less direct 
knowledge of the neutron density.  At present there is a large extrapolation
of experimental knowledge to the nearly pure neutron matter 
found between nuclei in the inner crust of neutron stars.

One avenue to extract additional information is to use microscopic
theories of neutron matter in an external field.  Different external
fields can be used to probe different properties of neutron matter;
to date finite systems have been studied using harmonic oscillator
or Woods-Saxon external one-body potentials.  These serve to confine
the system and to constrain the densities.  Originally these calculations
were limited to very small systems~\cite{Pudliner:1996}, but more recently
much larger systems have been treated where comparisons to density functionals
are more meaningful~\cite{Gandolfi:2010}.

Other properties of density functionals can also be studied
in the pure neutron limit.  The relative importance of
pairing and shell gaps is tightly constrained by the measured properties
of nuclei.  Shell gaps in neutron matter are expected to be much smaller
because of the strong pairing.  Spin-orbit splitting can also be
examined by studying systems away from closed shells.

\subsection{Neutron Matter and Density Functionals}

 It is again instructive to compare results for confined systems
in cold atoms with those in neutron matter.  Cold atoms at unitary
have essentially no closed shells, as discussed below, and hence can be
described as an expansion around the local density:
\begin{equation}
{\cal E}_2 \ = \ \int V_{ext} (r) \rho (r) \ + \xi \frac{3}{5}
( 3 \pi^2 )^{2/3} \rho^{5/3} \ + c_2 \ \nabla \rho^{1/2} \cdot \nabla \rho^{1/2}
+ \ c_4
\frac{\nabla^2 \rho^{1/2} \nabla^2 \rho^{1/2}}{\rho^{2/3}}  + ...,
\label{eq:densitygradient}
\end{equation}
where the first two terms provide the simple local-density approximation (LDA).

The form of this energy density functional is severely restricted by the scale-invariance
of the cold atom system.  All terms in the energy
density must scale like $\rho^{5/3}$.  Such a functional will not work
for cold atoms at small $k_F a$ as in this BCS regime individual particles
can propagate over the entire system and 
the fermionic closed shells are very important. While it is not
obvious $\textit{a priori}$ that this simple density functional will
work at unitarity, microscopic results using both Diffusion Monte
Carlo and lattice Auxiliary Field Monte Carlo justify this
simple local density plus gradient expansion form~\cite{Carlson:2014b}.

 Clearly such a simple form will not work well for nuclei or neutron drops,
where the fermionic shell closures play a critical role.  Of course
Skyrme and Hartree-Fock-Bogoliubov models play a critical role in the
study of medium and heavy nuclei, including particularly neutron-rich 
nuclei~\cite{bender2003self}. These density functionals have parameters
that are fit to the properties of nuclei, and sometimes to the calculated
bulk properties of neutron matter. A similar strategy should be helpful
in constraining the gradient, spin-orbit, and pairing terms in very
neutron-rich nuclei.

\subsection{Neutron Drops}

\begin{figure}[!tb]
\centering
   \begin{subfigure}{0.48\textwidth} 
     \includegraphics[scale=0.22]{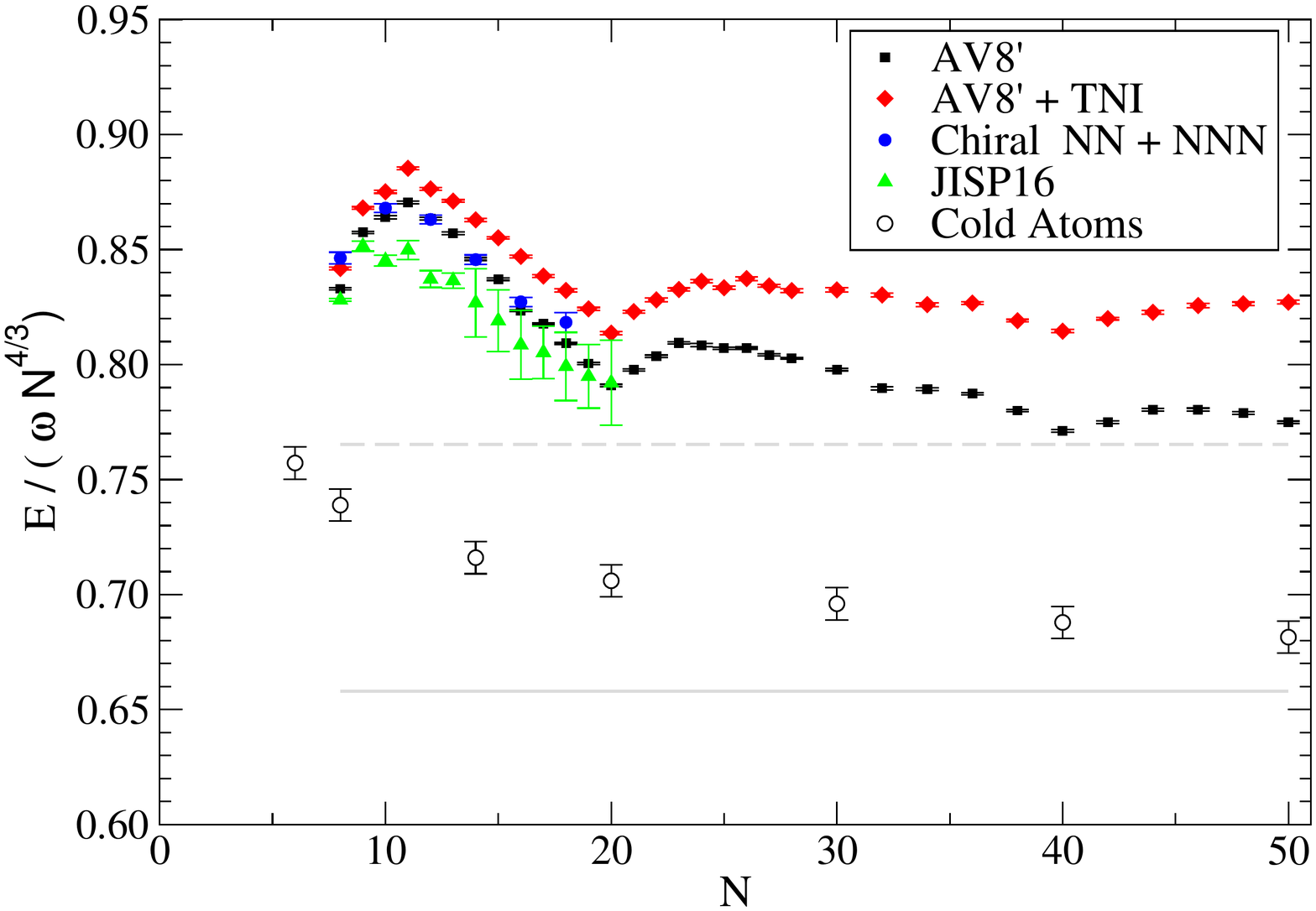}
     \caption{}
   \end{subfigure}
   \begin{subfigure}{0.48\textwidth}
     \includegraphics[scale=0.26]{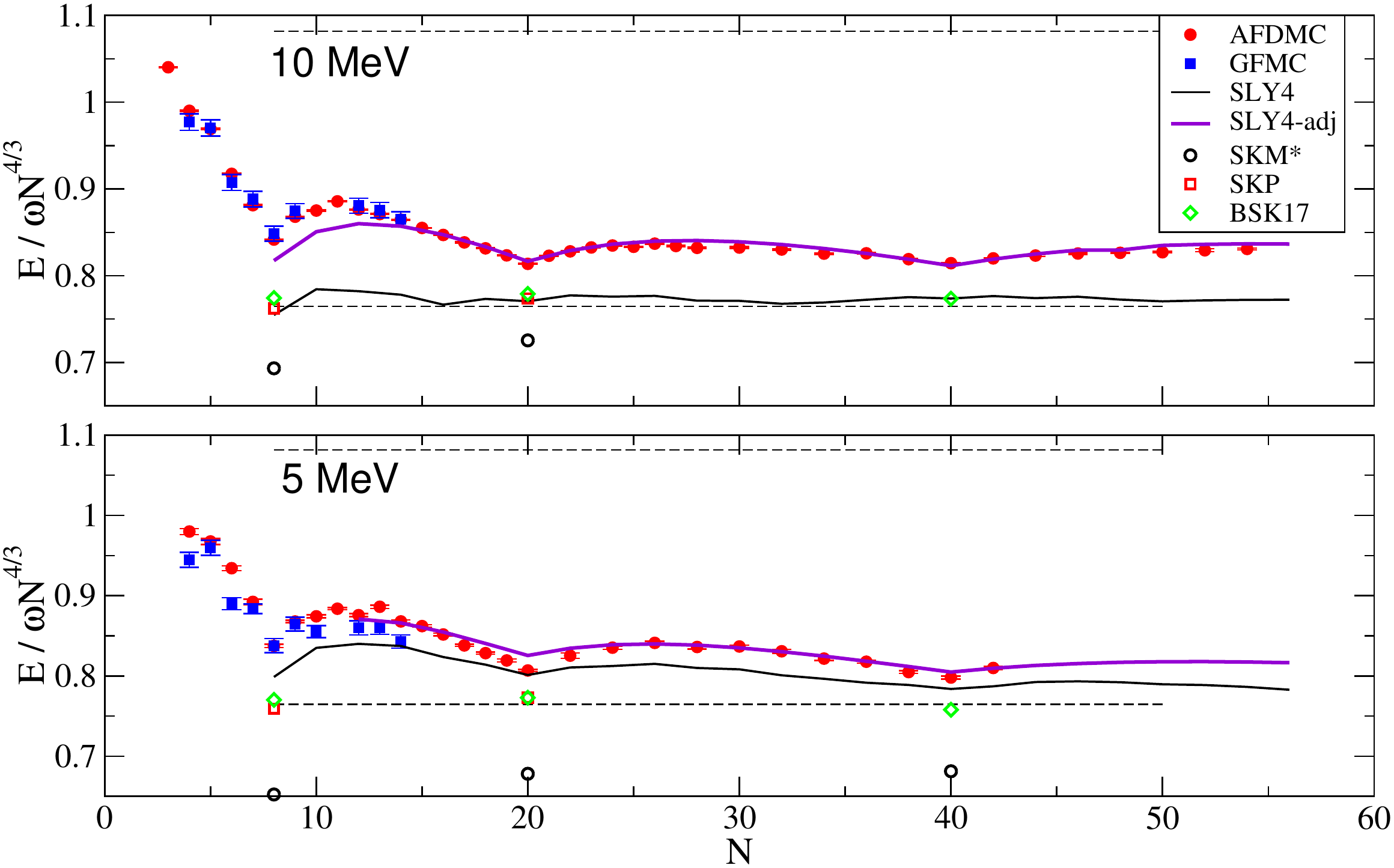}
     \caption{}
   \end{subfigure}
  \vspace{0.5cm}
\caption{Energies for neutrons trapped in a harmonic oscillator with $\omega = 10$ MeV for different interactions and methods compared to Monte Carlo calculations for cold atoms at unitarity (open symbols) and two local density approximations (solid and dashed lines).  In the right panel results for the AV8' + UIX interaction are compared to several pre-existing density functionals.}
\label{fig:ndropsho}
\end{figure}

  Here we summarize results for neutron drops in Hamonic
Oscillator and Wood-Saxon external fields, and compare them to
available results for cold atoms. A variety of different interactions
and methods have been used for calculations of neutron 
drops~\cite{Gandolfi:2010,Maris:2013drops,Potter:2014drops,Bogner:2011}.
In Fig. \ref{fig:ndropsho} we show results for neutrons trapped in a
harmonic oscillator trap with a frequency of 10 MeV.  Results for
lower and higher trap frequencies have also been reported.
For cold atoms lattice methods are used which are free of a sign
problem; note that the cold atom results show essentially no evidence
of the closed shells that would appear for non-interacting fermions with 
N=8, 20, 40, etc.  The local density approximation with 
the bulk value of $\xi = 0.37$ for cold atoms
is shown as a solid black line, finite drops rapidly and smoothly
approach this bulk limit.  Gradient corrections have been studied in
Ref.~\cite{Carlson:2014b}.  The dashed line indicates the
local density approximation for $\xi = 0.50$, a value more appropriate
for the neutron equation of state.

Calculations from 8 to 50 neutrons using the AV8' NN interaction
and the AV8' + UIX NNN interaction are shown. The results plotted
here use the AFDMC method; results for N up to 16 with GFMC are essentially
identical.  The three-nucleon interaction is not very important for small
N, as the density is relatively modest. For larger N, though, the central
density in a harmonic trap can be quite large and the three-nucleon interaction
makes a considerable difference.  Results are also shown for no-core shell
model (NCSM) calculations using chiral interactions and the JISP16 NN
interaction. The chiral interactions are very similar to the AV8' results
without a three-nucleon interaction; the energies for all interactions
and methods agree within a few percent.

These calculations show significant shell closures, in contrast to the
cold atom results.  These shell closures arise because of the effective
range in the neutron-neutron interaction, and the concominant smaller 
size of the superfluid pairing
gap.  In cold atoms only superfluid pairs propagate across
the whole system, while for neutrons the shell closures indicate the
single-particle picture still survives to some degree. The shell closures
are not as strong as in nuclei, however.  Pairing gaps are also evident
from the odd-even staggering, spin-orbit, and gradient corrections have
been examined in many of these studies~\cite{Maris:2013drops,Potter:2014drops}.

In the right-hand panel, results for the AV8'+UIX interaction are compared
with a variety of previously existing density functionals.  These
funtionals do not fully describe the neutron drop results, in general
their energy is too low and spin-orbit and pairing are not completely correct.
Modest modifications to the isovector (full neutron) gradients, pairing and
spin orbit give density functional results in the full black line, labeled
SLY4-adj.  These do a good job of reproducing results in both 5 and 10 MeV
frequency traps except for the smallest systems considered.
More recently it has been shown that new density functionals, i.e.
UNEDF0~\cite{Kortelainen:2010}  and UNEDF1~\cite{Kortelainen:2012}, can be created
that simultaneously fit nuclei with accuracy comparable or better than
the existing functionals, and also reproduce the neutron drop results.

\begin{figure}[!tb]
\centering
   \begin{subfigure}{0.48\linewidth} 
     \includegraphics[scale=0.26]{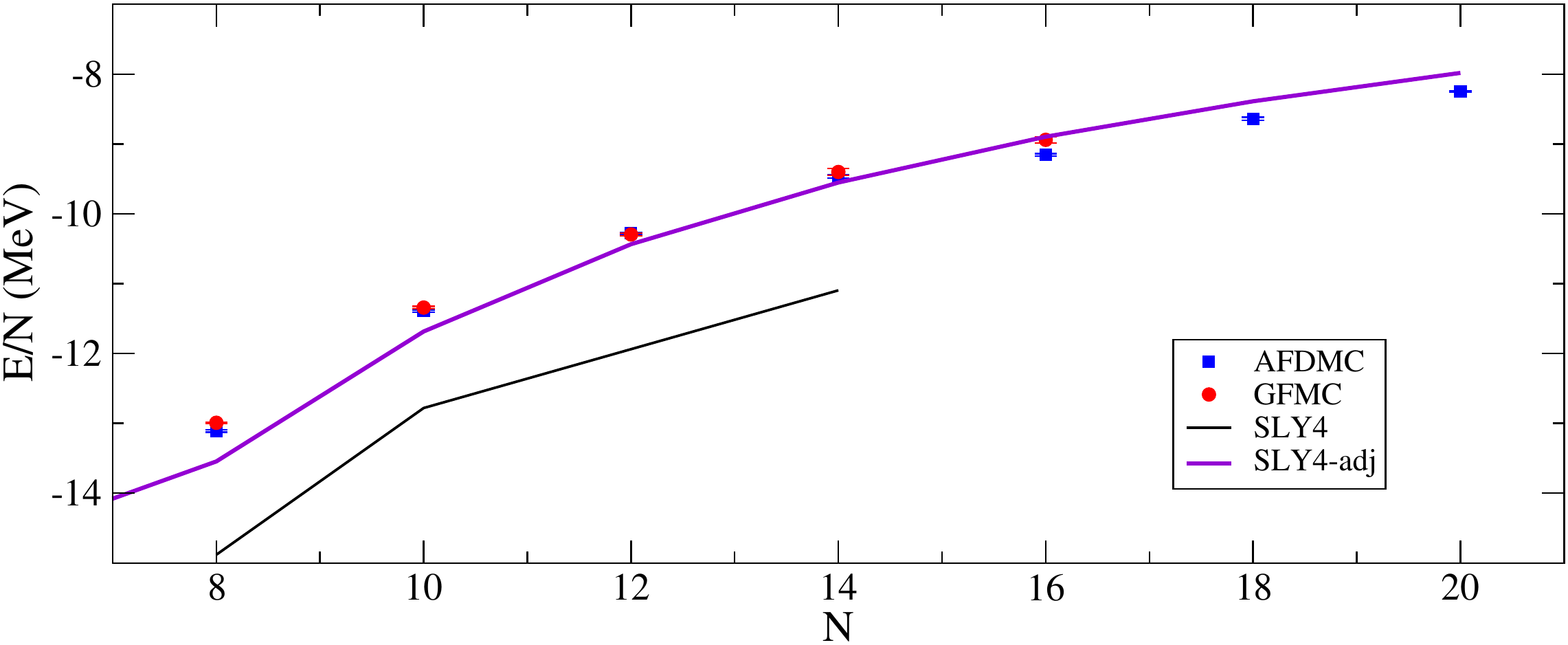}
     \caption{}
   \end{subfigure}
   \begin{subfigure}{0.48\textwidth}
     \includegraphics[scale=0.22]{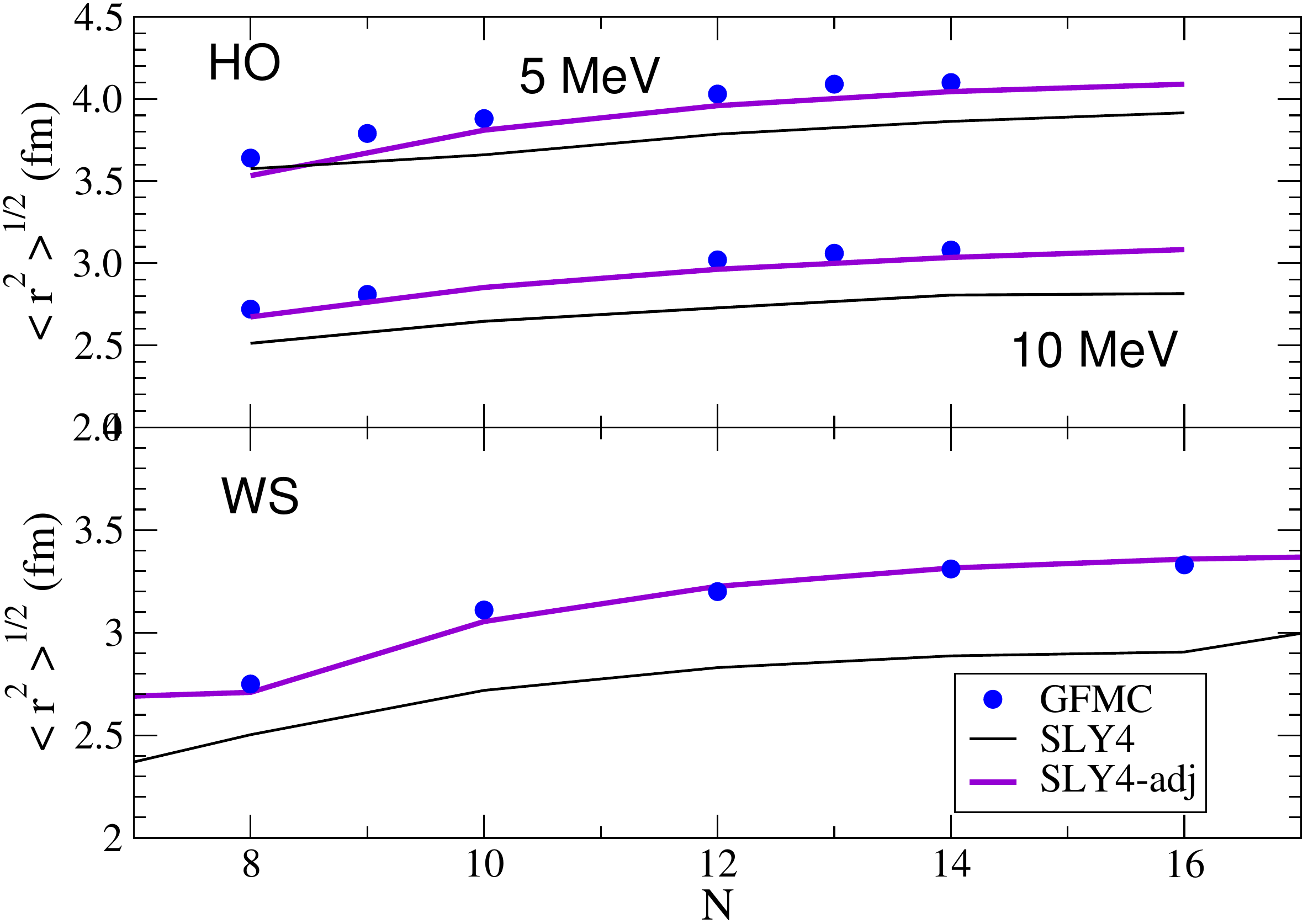}
     \caption{}
   \end{subfigure}
  \vspace{0.5cm}
\caption{Energies for neutrons trapped in a Wood-Saxon external field compared with density functionals (left panel) and radii of neutron drops in HO and WS wells compared to density functionals (right panel).}
\label{fig:dropws}
\end{figure}

In figure \ref{fig:dropws} we show results in a Wood-Saxon well.
These more nearly mimic nuclei since the central density is rather
fixed, and hence they are not so sensitive to the high-density behavior
of the equation of state. The same modified density functional that fits
the harmonic oscillator results also reproduces the Wood-Saxon results,
and simultaneously describes the calculated radii of drops in both
external fields.

Further studies of inhomogeneous neutron matter are warranted, including
effects of the three-nucleon interaction and different external fields.
In the inner crust of a neutron star, for example, the neutrons are
not small finite systems. Therefore it would be very useful to constrain
the gradient terms with periodic external fields with a range of momenta,
and see if the same density functional can describe these systems as well
as finite drops.

\section{HIGHER DENSITIES: NEUTRON MATTER AND NEUTRON STARS}
At higher densities the three-nucleon interaction becomes
even more important.  These play an important role in the mass/radius
relation for neutron stars. In this section we describe
the equations of state obtained at higher densities including
three-nucleon interaction results and the equation of state, 
describe the relation between the symmetry energy $E_{\rm sym}$, the difference
between neutron matter energy and isospin-symmetric nuclear matter
energy at nuclear saturation density, and its density derivative $L$.
Finally we describe the impact of these constraints on the neutron star
mass-radius relation as well as describe  briefly the importance
of three-hadron neutron-neutron-lambda interactions in the equation
of state of high-density neutron matter.

\subsection{Equation of State}
The EOS of neutron matter around nuclear densities is important for several
reasons. As outlined in the previous sections, in this regime several channels
contributing to the nucleon-nucleon interaction are important, and thus the
EOS provides a direct comparison between different frameworks for the nuclear
Hamiltonian. 
Although the maximum mass of neutron stars is dominated by the EOS
at very high densities, their radius is determined by the pressure in the 
region of about 1 to 2 $\rho_0$, and then measurements of radii of neutron 
stars can be used to constrain the EOS.

\begin{figure}[!tb]
\centering
   \begin{subfigure}{0.49\linewidth} 
     \includegraphics[scale=0.26]{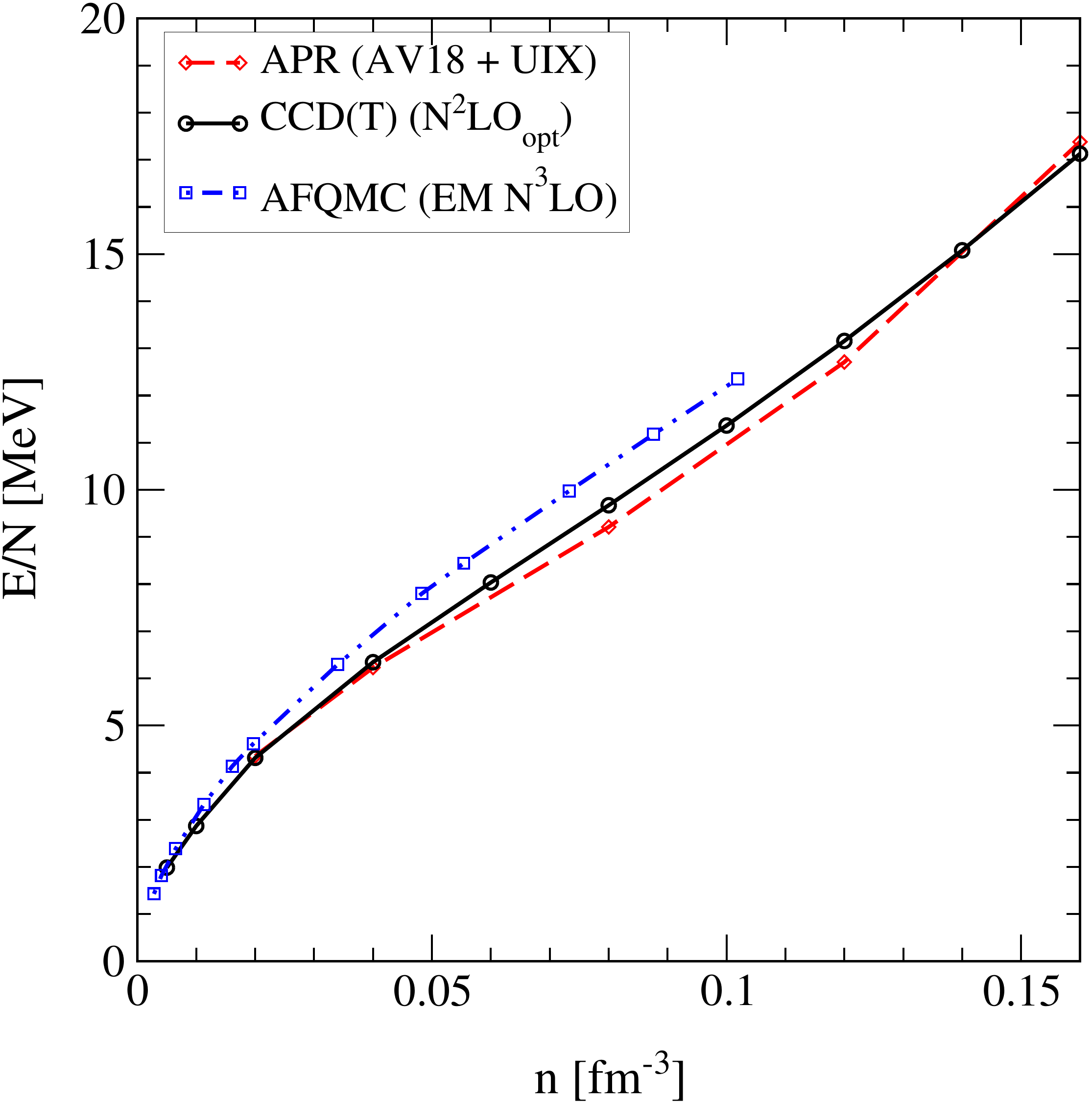}
     \caption{}
   \end{subfigure}
   \begin{subfigure}{0.49\textwidth}
     \includegraphics[scale=0.26]{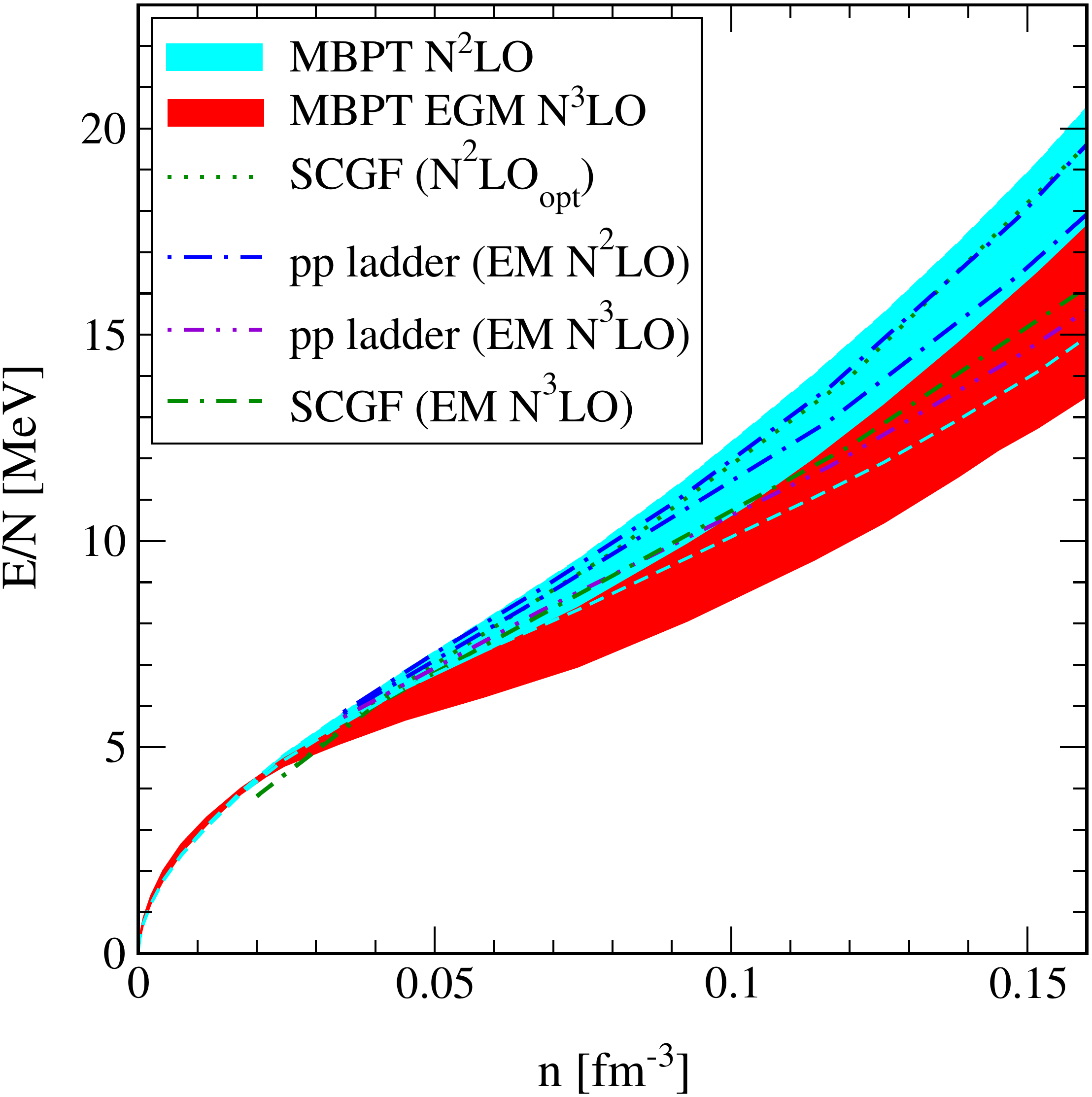}
     \caption{}
   \end{subfigure}
  \vspace{0.5cm}
\caption{Equation-of-state of intermediate density neutron matter using both NN interactions and 3NF
as input. Shown are results for different chiral EFT potentials, different orders in the chiral expansion, and different quantum many-body methods. (a) The standard APR results, along
with Coupled Cluster and Auxiliary-Field Quantum Monte Carlo results (b) Many-body perturbation
theory, Self-Consistent Green's Function, and pp ladder approximation results.
Details on the interactions employed are given in the main text.
}
\label{fig:eos_3b}
\end{figure}

We show in Fig.~\ref{fig:eos_3b} results for the equation of state of neutron matter
at somewhat smaller densities, following from 
calculations that use both nucleon-nucleon and three-nucleon interactions as input. 
Specifically, we show we show MBPT results
using EM and EGM N$^2$LO interactions as input (the lower end of this band is shown using 
a dashed line), \cite{Tews:2013} MBPT results using 
N$^3$LO EM potentials, \cite{Tews:2013}
results following from a particle-particle ladder approximation using N$^2$LO EM potentials, \cite{Sammarruca:2014}
as well as pp ladder results using an N$^2$LO EM potential, \cite{Sammarruca:2014}
self-consistent Green's functions (SCGF) results using N$^2$LO$_{\text{opt}}$, \cite{Carbone:2014}
as well as SCGF results using EM N$^3$LO as input, \cite{Carbone:2014} 
Coupled-Cluster with doubles and including some triples effects, CCD(T), results using the N$^2$LO$_{\text{opt}}$ interaction, \cite{Hagen:2014}
Auxiliary-Field Quantum Monte Carlo (AFQMC) results using an N$^3$LO EM potential, \cite{Wlazlowski:2014}
as well as the frequently cited Akmal-Pandharipande-Ravenhall results using AV18 plus UIX \cite{Akmal:1998}.

The most obvious feature of Fig.~\ref{fig:eos_3b} in comparison to Fig.~\ref{fig:eos_2b} is that three-nucleon 
interactions in neutron matter are repulsive, i.e. the energy goes up: 
this is in contradistinction with what holds in light and medium-mass nuclei. Another general
aspect of this figure relates to the transition from N$^2$LO to N$^3$LO: in Fig.~\ref{fig:eos_3b}b for MBPT, SCGF and
the pp ladder approximation results are shown for both orders in the chiral expansion, and in both cases the energy
decreases. This is despite the fact that the MBPT results use consistent interactions, i.e. 
include an N$^3$LO 3N force along
with the N$^3$LO NN force, while the pp ladder and SCGF values, like most other many-body results currently available, 
combine the N$^3$LO NN force with N$^2$LO at the 3N level. 
Also prominent is the fact that the SCGF N$^2$LO$_{\text{opt}}$ results reasonably match
the EM N$^2$LO pp ladder values (and similarly for the corresponding EM N$^3$LO ones), 
which is probably not unexpected (with the exception of the low-density region). 
A final noteworthy feature is that
the CCD(T) results appear to match rather nicely with the APR values. 
A conclusion that the Figs.~\ref{fig:eos_2b} \& \ref{fig:eos_3b} really underline 
is the significance of producing error bands using more than one potential as input, 
if possible employing a non-perturbative or otherwise controlled approximation scheme.

Around density $\rho_0$, the EOS of pure neutron matter is directly related
to the symmetry energy and its slope, and the analogous (and much more 
challenging) calculation of symmetric nuclear matter is not needed. 
The symmetry energy $E_{\rm sym}(\rho)$ is generally defined as the difference
between the energy of pure neutron matter and the energy of symmetric 
nuclear matter with the same total barion density.
In terms of the isospin asymmetry, $x\equiv\rho_p/\rho$, the energy per 
nucleon of isospin asymmetric nuclear matter is commonly expanded in even 
powers of $x$,
\begin{equation}
E(\rho,x)=E_0(\rho)+E_{\rm sym}^{(2)}(\rho)(1-2x)^2+
E_{\rm sym}^{(4)}(1-2x)^4+\dots \,,
\end{equation}
where $E$ is the energy per nucleon of the system,
$E_0(\rho)=E(\rho,x=0.5)$ is the EOS of symmetric
nuclear matter, and $E_{\rm sym}^{(4)}(\rho)$ and higher order
corrections are ignored here.
The symmetry energy $E_{\rm sym}$ is then given by
\begin{equation}
E_{\rm sym}(\rho)=E(\rho,0)-E_0(\rho) \,,
\end{equation}
where $E(\rho,0)$ is the EOS of pure neutron matter.
Near the nuclear saturation density, $\rho_0$, there are a number of
constraints on the EOS of infinite nuclear matter from
nuclear masses, charge radii, and giant resonances.
The extrapolation of the binding energy of heavy nuclei to the thermodynamic 
limit yields $E_0(\rho_0)=-16.0\pm0.1$ MeV~\cite{Moller:1995}. Because the 
pressure is zero at saturation, the symmetry energy can be expanded around 
saturation as
\begin{equation}
E_{\rm sym}(\rho)\Big|_{\rho_0}= E_{\rm sym}+
\frac{L}{3}\frac{\rho-\rho_0}{\rho_0}+\dots \,,
\label{eq:lvsesym}
\end{equation}
where $E_{\rm sym}$ is the symmetry energy at saturation, and $L$ is a parameter 
related to its slope.
It is convenient to first focus to $E_{\rm sym}$ at $\rho_0$ for several reasons.
The calculation of symmetric nuclear matter is much more challenging than
pure neutron matter, mainly because the Hamiltonian is more complicated,
and isospin
symmetric matter is more strongly correlated through the tensor
interaction acting in $S$- and $D$-waves.
In particular, in the low density regions clustering effects
are very difficult to be included in the calculation of the EOS, and/or 
isolated from experimental measurements. 

Using QMC methods, the EOS of pure neutron matter is typically calculated
with a simulation of 66 particles in a box imposing periodic
boundary conditions, for which finite size effects are generally very well
under control~\cite{Sarsa:2003,Gandolfi:2009}. 
A comprehensive study of the model of the three-neutron interaction has been
presented in Ref.~\cite{Gandolfi:2012}, with particular emphases to the role 
played by short range correlations.
Within the Urbana/Illinois models, the main contribution of the three-body force is given
by the short-range part, whose structure in the UIX model is
\begin{equation}
\label{eq:vr}
V_{ijk}^R=A_R \sum_{\rm cyc} T^2(m_\pi r_{ij})T^2(m_\pi r_{jk}) \,,
\end{equation}
where $m_\pi$ is the pion mass, and
\begin{equation}
T(x)=\left(1+\frac{3}{x}+\frac{3}{x^2}\right)\frac{e^{-x}}{x}\xi^2(x) \,.
\label{eq:tx}
\end{equation}
The function $\xi(x)=1-e^{-cx^2}$ is a cutoff function to regularize
$T(x)$ at small distances. The $V^R$ term basically models the
four-pion exchange between neutrons~\cite{Carlson:1983}. In order to address the
role of the above short-distance behavior, other forms for $V_{ijk}^R$
have been considered, in addition to different models of intermediate-
and long-range contributions of the three-body force. These terms, where
two or three pions are exchanged between neutrons, with the creation of
$\Delta$ excited states, are described in Ref.~\cite{Pieper:2001}.

\begin{figure}[!tb]
\begin{center}
\includegraphics[width=0.7\columnwidth]{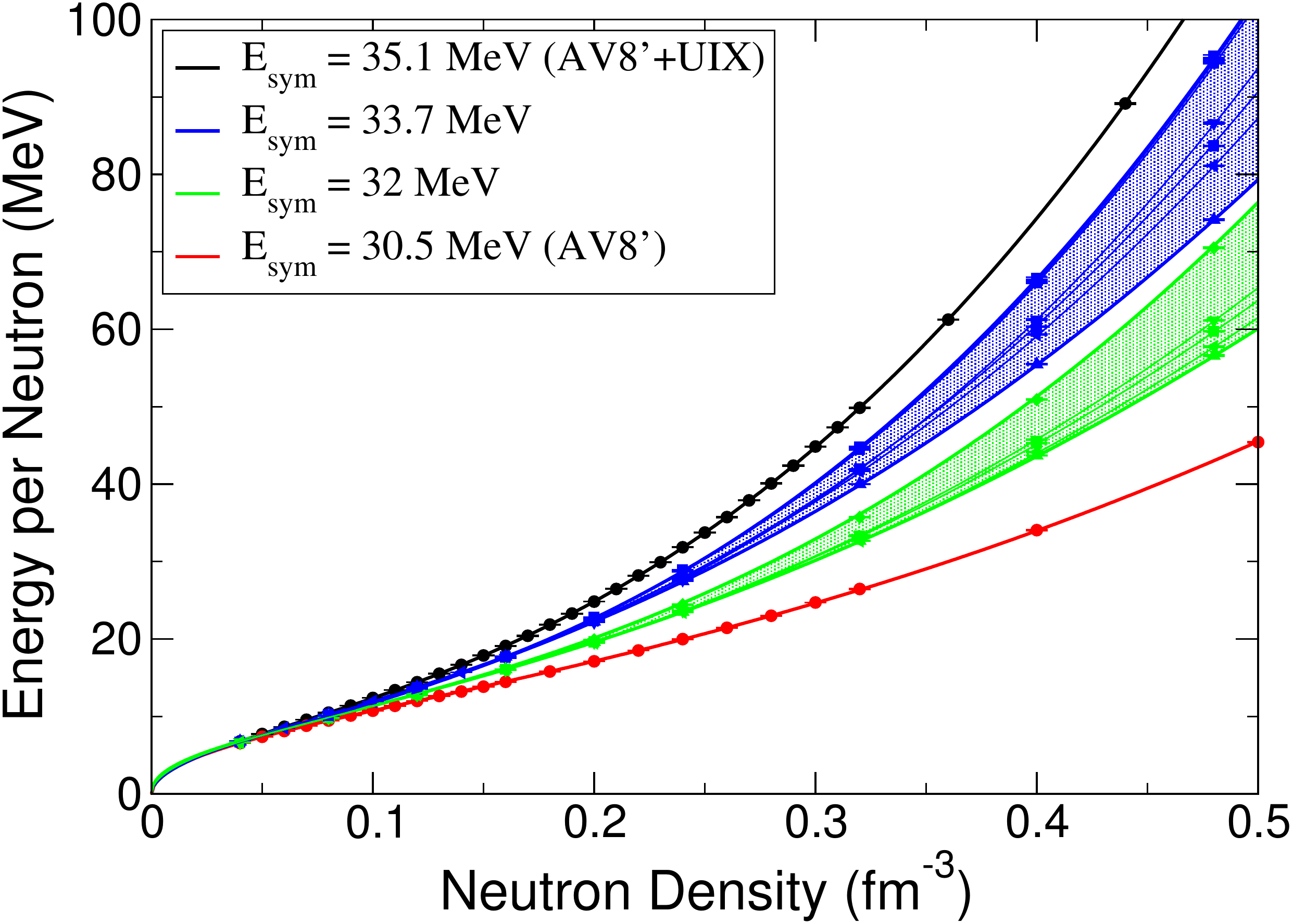}
\end{center}
\caption{The equation of state of neutron matter obtained by
using various models of three-neutron force as described in the 
text. For each model we impose that the energy at saturation is
17.7(1) MeV (blue band), or 16.0(1) MeV (green band). The results
are compared with the equations of state obtained with the AV8$'$ and
AV8$'$+UIX Hamiltonians. In the legend we indicate the corresponding
symmetry energy at saturation. 
Figure taken from \cite{Gandolfi:2014}.
}
\label{fig:eosall}
\end{figure}

The resulting EOS are shown in Fig.~\ref{fig:eosall}, where we compare the 
results obtained with the AV8$'$ and AV8$'$+UIX Hamiltonians with several EOS
obtained using different models of three-neutron interactions, adjusted to
give the value for $E_{\rm sym}$ indicated in the figure.
The range of $E_{\rm sym}$ is compatible with several experimental 
measurements~\cite{Tsang:2012,Lattimer:2012}.
To obtain a lower symmetry energy with the AV8$'$ model of two-body force
would require an attractive contribution of the three-body force.
The three-pion rings give attraction in pure neutron matter~\cite{Maris:2013}, but they
would require a strong short-range repulsion to make the EOS stiff enough
to support astrophysical observations.

The EOS calculated using QMC can be conveniently parametrized using the 
following functional form:
\begin{equation}
E(\rho_n)=a\left(\frac{\rho_n}{\rho_0}\right)^\alpha+
b\left(\frac{\rho_n}{\rho_0}\right)^\beta \,,
\label{eq:fit}
\end{equation}
where $E(\rho_n)$ is the energy per neutron as a function of the 
neutron density $\rho_n$, and the parameters $a$, $\alpha$, $b$, and $\beta$
are obtained by fitting the QMC results. 
The parametrization of the equations of state obtained with the AV8$'$ and 
AV8$'$+UIX Hamiltonians are reported in~\cite{Gandolfi:2014}.

\begin{figure}[!tb]
\begin{center}
\includegraphics[width=0.7\columnwidth]{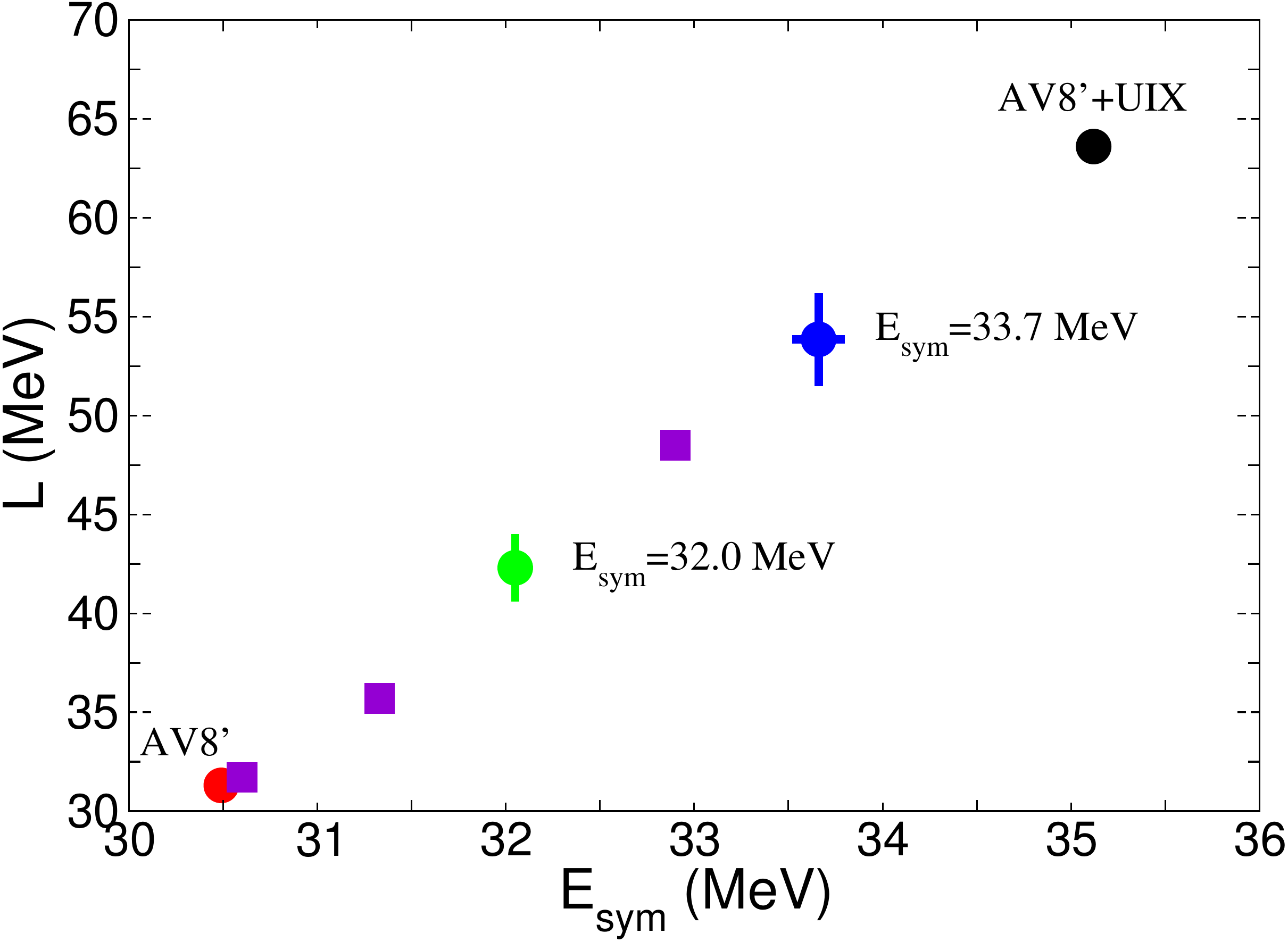}
\end{center}
\caption{The value of L as a function of E$_{\rm sym}$ obtained from
various EOS. The green and blue points with error bars correspond to
the various EOS indicated by the two colored areas of
Fig.~\ref{fig:eosall}, and red and black points show the results
obtained using a two-body force alone and combined with the UIX
model. The square symbols correspond to results obtained by
independently changing the cutoff parameters entering in $V_R$ and
in the three-pion rings of the three-neutron force.
Figure taken from \cite{Gandolfi:2014}.
}
\label{fig:lvsesym}
\end{figure}

Using Eq.~(\ref{eq:lvsesym}) the value of $E_{\rm sym}$ and L can be easily
extracted from the calculated EOS. 
The result are shown 
in Fig.~\ref{fig:lvsesym}, where we compare the results
obtained using the AV8$'$ and AV8$'$+UIX Hamiltonians (red and black
symbols), the various EOS giving the indicated $E_{\rm sym}$ obtained by
changing the three-neutron force model (green and blue symbols), and
results obtained using the Illinois model of three-neutron force that
includes three-pion rings where we have independently changed the
cutoff of the intermediate- and short-range part. It is clear that
within this model the correlation between L and $E_{\rm sym}$ is very
strong.

\subsection{Neutron Star Mass/Radius}
The neutron star matter is mainly composed by neutrons and a few protons. 
 When the EOS of the neutron-star matter has been specified, the structure
of an idealized spherically-symmetric neutron star model can be calculated
by integrating the Tolman-Oppenheimer-Volkoff (TOV) equations:
\begin{equation}
\frac{dP}{dr}=-\frac{G[m(r)+4\pi r^3P/c^2][\epsilon+P/c^2]}{r[r-2Gm(r)/c^2]} \,,
\qquad
\frac{dm(r)}{dr}=4\pi\epsilon r^2 \,,
\label{eq:tov2}
\end{equation}
where $m(r)$ is the gravitational mass enclosed 
within a radius $r$, and $G$ is the gravitational constant.
The above equations are solved by obtaining the energy density $\epsilon$ and
pressure $P$ from the EOS, and by specifying an initial point in the
integration that is given by the central density of the star. For the
EOS with the form of Eq.~\ref{eq:fit} we have
\begin{equation}
\epsilon=\rho_0\left[a\left(\frac{\rho}{\rho_0}\right)^{1+\alpha} 
+b\left(\frac{\rho}{\rho_0}\right)^{1+\beta}+m_n\left(\frac{\rho}
{\rho_0}\right)\right]\,,
\end{equation}
and
\begin{equation} 
P=\rho_0\left[a\alpha\left(\frac{\rho}{\rho_0}\right)^{1+\alpha}
+b\beta\left(\frac{\rho}{\rho_0}\right)^{1+\beta}\right]\,.
\end{equation}
The solution of the TOV equations gives, for a specified central density $\rho_c$,
the profiles of $\rho$, $\epsilon$ and $P$
as functions of radius $r$, and also the total radius $R$ and mass
$M=m(R)$. The total radius $R$ is defined to the point where the 
pressure vanishes.
The solution of the TOV equations are modified only slightly by magnetic fields
and temperatures which are expected, and rotation is less than a
10\% effect for the kinds of M-R curves which we present below.
The speed of
sound, $c_s$ in the neutron star interior is $c_s^2 =
dP/d\epsilon$, and ensuring that this is less than the speed of
light (and thus the EOS is said to be ``causal'') constrains the
set of possible EOS. Also, the pressure must increase with
increasing energy density, $dP/d\epsilon>0$, in order to ensure
that the neutron star is hydrodynamically stable.

\begin{figure}[!tb]
\begin{center}
\includegraphics[width=0.7\columnwidth]{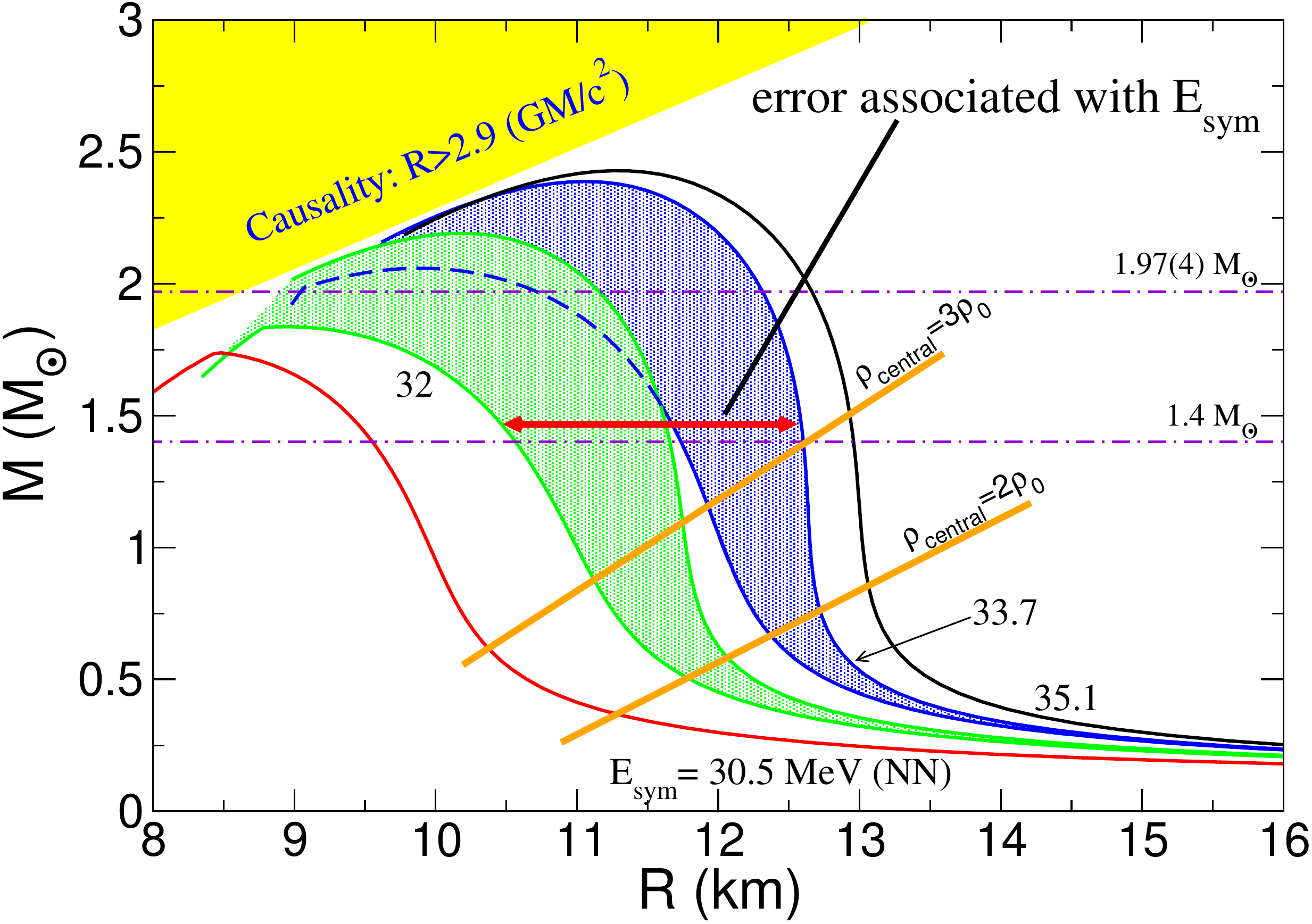}
\end{center}
\caption{The mass-radius relation for neutron stars based on the QMC
neutron matter results above. Results are presented for the
different EOS given in Fig.~\ref{fig:eosall}. The numbers indicate
the value of $E_{\rm sym}$ of the various EOS.
}
\label{fig:mr}
\end{figure}

In practice, the crust of neutron stars cannot be modeled with 
a pure neutron matter EOS. Several approaches to describing 
the EOS of neutron-star matter exist, including the combination of
microscopic calculations based on chiral EFT  
with polytropes put forward in \cite{Hebeler:2010b,Hebeler:2013}.
All the results presented in this section
have been obtained by using QMC calculations shown in the above
sections for densities $\rho \ge \rho_{\rm crust}=0.08$ fm$^{-3}$. 
At lower densities the neutron star matter is modeled by 
using the EOS of the crust obtained in Refs.~\cite{Baym:1971,Negele:1973}. 
When the EOS violates causality (this is very often the case of 
EOS obtained from non-relativistic nuclear Hamiltonians), the EOS
is switched to the maximally stiff EOS.
The TOV equations are solved for several values of $\rho_c$, and the 
solution giving the maximum mass is considered.

By using the EOS obtained from different nuclear Hamiltonian, we
can study the effect to the neutron star structure.
The results of the M-R diagram of neutron stars obtained from the 
EOS calculated in the previous section are shown in Fig.~\ref{fig:mr}. 
Since the radii of neutron stars are almost determined by the EOS 
slightly above $\rho_0$~\cite{Lattimer:2001}, future measurements will
provide strong constraints to the nuclear Hamiltonian.
In particular, radii are directly connected to the pressure of 
neutron matter at $\rho_0$, and then there is a natural 
correlation between $E_{\rm sym}$ and $L$ and radii.
In the figure the two bands correspond to the EOS described in
the previous section (the corresponding values of the symmetry 
energy are also indicated in the figure). The red and black 
curves correspond to the EOS calculated with the AV8$'$ two-body 
interaction alone, and combined with the UIX three-neutron potential.
The relation between $E_{\rm sym}$ and the radius is evident,
as the increasing of $E_{\rm sym}$ predicts a neutron star with
a larger radius.
In the figure, the density of the neutron matter inside the star 
is indicated with the orange lines. As anticipated, even at large masses
the radius of the neutron star is mainly governed by the equation of
state of neutron matter between 1 and 2 $\rho_0$~\cite{Lattimer:2001}.

As is clear from the figure, the AV8$'$ Hamiltonian alone does not 
support the recent observed neutron star with a mass of 
1.97(4)M$_\odot$~\cite{Demorest:2010} and 2.01(4)M$_\odot$ \cite{Antoniadis:2013}.
the addition of a three-body force to AV8$'$ can provide sufficient
repulsion to be consistent with all of the
constraints. 
The results also suggest that the most modern neutron matter EOS
imply a maximum neutron star radius not larger than 13.5 km, unless
a drastic repulsion sets in just above the saturation density~\cite{Gandolfi:2012}. This
rules out EOS with large values of $L$, typical of Walecka-type
mean-field models without higher-order meson couplings which can
decrease $L$.
We note that our analysis suggests it is unlikely that neutron stars 
have radii lower than 10 km~\cite{Ozel:2010,Guillot:2013}, and/or larger than 15 km,
but determining the systematic uncertainties is still an open question~\cite{Heinke:2014}.

These recent observations of neutron stars with
1.97(4)M$_\odot$~\cite{Demorest:2010} and 2.01(4)M$_\odot$
\cite{Antoniadis:2013} put the most severe constraints on the EOS,
although the precise hadronic composition is still undetermined.
When the EOS is stiff, the corresponding 
chemical potential is large enough that heavier particles, hyperons,
can be created from neutrons. For example, in the case of 
neutrons the $\Lambda$ particles might form, and a fundamental 
question is about their role to the EOS. Note that other 
lighter hyperons (like $\Sigma$) would form at higher densities 
because they need that the proton fraction is large enough, and 
consequently their production is suppressed compared to $\Lambda$.
The main uncertainty in the inclusion of hyperons is because
the nucleon-hyperon (and the hyperon-hyperon) interaction is 
not well understood.
For example, several calculations based on Brueckner-Hartree-Fock
suggest that when hyperons are included in the EOS, the corresponding
mass of neutron stars is very low~\cite{Schulze:2011}. Other calculations based
on relativistic mean-field or similar methods instead do not 
exclude that the inclusion of hyperons give neutron stars
that are not compatible with recent observations.
The $\Lambda$-nucleon potential, and in particular the importance of
the $\Lambda$-neutron-neutron interaction, has been recently discussed
by comparing QMC calculations of the binding energy of hypernuclei with
available experimental measurements~\cite{Lonardoni:2013,Lonardoni:2014}.
The role of $\Lambda$ hyperons in neutron star matter has been discussed
by Lonardoni et al. in Ref.~\cite{Lonardoni:2014b}. The main conclusion is
that a small change of the $\Lambda$-neutron-neutron interaction, while
producing very small changes to the properties of hypernuclei, has a
dramatic effect in neutron matter.

\begin{figure}[!tb]
\begin{center}
\includegraphics[width=0.7\columnwidth]{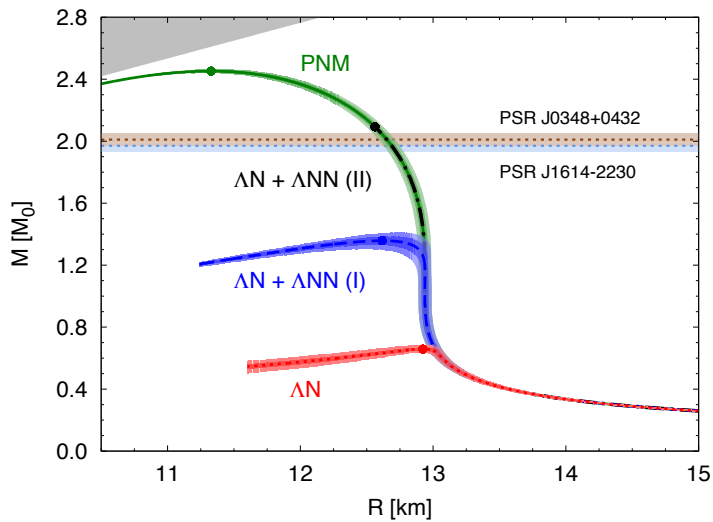}
\end{center}
\caption{The mass-radius for neutron stars obtained from EOS that
include $\Lambda$-hyperons. The green upper line (denoted with PNM) 
is the result of pure neutron matter (the same presented in Fig.~\ref{fig:mr}).
The red bottom band (indicated with $\Lambda$N) is obtained by including
only the $\Lambda$-nucleon interaction in the Hamiltonian.
The blue curve, $\Lambda$N+$\Lambda$NN (I), is obtained but including 
the Usmani three-body interaction~\cite{Usmani:1995}. The EOS obtained using the
$\Lambda$N+$\Lambda$NN (II) does not allow to $\Lambda$ hyperons to produce,
and is indicated by the dashed black line. Recent heavy-mass neutron stars
measurements are also indicated by horizontal lines.
Figure taken from \cite{Lonardoni:2014b}.}
\label{fig:mr_hyp}
\end{figure}

In Fig.~\ref{fig:mr_hyp} we show the mass-radius of neutron stars obtained
by including the $\Lambda$ hyperons to the EOS. As described in 
Ref.~\cite{Lonardoni:2014b}, the fraction of $\Lambda$ is determined 
by imposing chemical equilibrium once the energy as a function of the 
concentrations of $\Lambda$ is known. That has been calculated using
the AFDMC method for different Hamiltonians.
The AV8$'$+UIX for the pure neutron sector has been combined with 
the $\Lambda$-nucleon interaction of Ref.~\cite{Usmani:1995}, and with the addition
of two different $\Lambda$-nucleon-nucleon potentials, denoted as
$\Lambda$N+$\Lambda$NN (I) and $\Lambda$N+$\Lambda$NN (II).
While the operatorial structure of these forces is the same, the 
strengths have been slightly changed to reproduce better the 
binding energy of $\Lambda$-hypernuclei. Although the two Hamiltonians 
give qualitatively very similar results in hypernuclei, the EOS
and the corresponding neutron star structure are dramatically 
different.
It is clear that further input on hyperon-nucleon interactions, especially
on $\Lambda$NN, will be crucial if firm conclusions are to be drawn.

\section{CONCLUSIONS AND OUTLOOK}
  A great deal has been learned about neutron matter over the past decade,
both from theory and through experiments and observations.  The equation of
state of cold neutron matter at low and moderate densities is now tightly constrained from many-body calculations with realistic interactions.
Also, the pairing gap has been reliably extracted from methods that
correctly predicted the pairing gap in Fermi atoms at unitarity, a 
similarly strongly paired Fermi system.

  Calculations of the equation of state provide strong constraints on the
radii of typical neutron stars of masses $\approx$ 1.4 solar masses.
These seem to be in reasonable agreement with mass/radius constraints
extracted from astrophysical observations, but the observations and
particularly the associated extraction of the mass/radius 
relation remain a controversial area with a variety of results.

  Constraints on inhomogeneous matter are also starting to appear: 
 it is encouraging that these results can be made consistent with density functionals
that also describe nuclei.  Traditionally, density functionals have used limited
constraints at the extremes of isospin, so this remains a promising avenue
for future research.  It will be interesting to see if these constraints
alter predictions for the properties of neutron-star crusts.

  Finally, the observation of two-solar-mass neutron stars has challenged the
traditional view that hyperons and other non-nucleonic degrees of freedom
will appear and considerably soften the high-density equation of state.
In the case of hyperons, the role of the hyperon-two-nucleon interaction
is very important, and may limit the role of hyperons in dense neutron-rich
matter. Studies of cold dense matter from the fundamental degrees of freedom,
quarks and gluons, remain critical, though they are quite challenging.

  There are many prospects for important further advances, both in
theory and in experiments and observations.  Experiments at rare 
isotope facilities will
allow us to study many important neutron-rich nuclei, providing valuable
information on neutron-rich matter.  Observations and improved understanding
should enable better constraints 
on the neutron-star mass-radius relation.  The
observation of gravitational waves associated
with neutron-star mergers is an exciting possibility, and 
this could provide important constraints on neutron-star properties
in the near future.

  Our increasing ability to calculate properties of strongly interacting
quantum systems will also play an important role.  Neutron-star matter
typically contains approximately ten percent protons: the impact on the
neutron star mass/radius relation is likely to be modest but it may
be very important for dynamic properties including the weak response
of dense neutron matter.  The role of superfluidity at higher densities,
both p-wave superfluidity in neutrons and s-wave superfluidity of the
low-density protons, is another important problem.  Finally, studies of
matter at finite temperature are critical for both potential sites of
heavy-element synthesis, core-collapse supernovae, and neutron-star mergers.

%Disclosure
\section*{DISCLOSURE STATEMENT}
The authors are not aware of any affiliations, memberships, funding, or financial holdings that might be perceived as affecting the objectivity of this review.

% Acknowledgement
\section*{ACKNOWLEDGMENTS}
We would like to acknowledge valuable discussions with A. Carbone, E. Epelbaum, G. Hagen, 
K. Hebeler, J. W. Holt, T. Kr\"uger, A. Lovato,
D. Lonardoni, S. Pieper, S. Reddy, R. Sharma, K. Schmidt,
A. Schwenk, I. Tews, R. Wiringa, G. Wlazlowski.
The work of S.G. and J.C. was supported by the U.S.~Department of Energy,
Office of Nuclear Physics, under contract DE-AC02-05CH12231, and by the
NUCLEI SciDAC project. The work of A.G. was supported by the Natural Sciences and Engineering 
Research Council of Canada.
This work was accomplished with generous computer support provided
by the INCITE program, Argonne and Los Alamos National Labs, NERSC, and the J\"ulich Supercomputing Center.

\bibliographystyle{ar-style5}
\bibliography{nmatter}
\end{document}